%% file: ms.tex
\documentclass[journal,10pt, letterpaper]{IEEEtran}

\usepackage{booktabs}
\usepackage{psfrag}
\ifCLASSINFOpdf
   \usepackage[pdftex]{graphicx}

\else
  
   \usepackage[dvips]{graphicx}

\fi
\usepackage[cmex10]{amsmath}
\usepackage{amssymb}
\usepackage{amsthm,epstopdf,nicefrac,pgfplots}

\usepackage{float}

\usepackage{rotating}

\usepackage{mathtools}

\usepackage{gensymb}

\renewcommand\appendix{\par
  \setcounter{section}{0}
  \setcounter{subsection}{0}
  \setcounter{figure}{0}
  \setcounter{table}{0}
  \renewcommand\thesection{Appendix \Alph{section}}
  \renewcommand\thefigure{\Alph{section}\arabic{figure}}
  \renewcommand\thetable{\Alph{section}\arabic{table}}
}

\usepackage{color}

\usepackage{algpseudocode}
\usepackage{algorithm}

\newtheorem{theorem}{Theorem}

\twocolumn

  \newlength\fheight 
    \newlength\fwidth 
\usepackage{balance}
\usepackage{cite}
\usepackage{url}

\usepackage{algorithm}
\usepackage{algpseudocode}

\usepackage[tight]{subfigure}

\usepackage{circuitikz}

\usepackage{pgfplots}
\pgfplotsset{
tick label style={font=\footnotesize},
label style={font=\footnotesize},
legend style={font=\footnotesize},
compat=1.12
}

\usepackage{calc}

\allowdisplaybreaks[1]

\begin{document}
%
\title{Best Linear Approximation of Wiener Systems \\
Using Multilevel Signals: 
Theory and Experiments}
%
%
%


\author{
    A.~De Angelis, 
    J.~Schoukens, 
    K.~R.~Godfrey,
    P.~Carbone
\thanks{This work was supported in part by the Italian Ministry of Instruction, University and Research (MIUR) through grant PRIN 2015C37B25.}
\thanks{A.~De Angelis and P.~Carbone are with the Engineering Department,  University of Perugia, via G. Duranti 93, 06125 Perugia, Italy.
\{alessio.deangelis, paolo.carbone\}@unipg.it}
\thanks{J. Schoukens is with Department ELEC, Vrije Universiteit Brussel, Pleinlaan 2, B1050 Brussels, Belgium.}
\thanks{K.~R.~Godfrey is with the School of Engineering, University of Warwick, CV4 7AL, Coventry, United Kingdom.}

\thanks{\copyright 2017 IEEE. Personal use of this material is permitted. Permission from IEEE must be obtained for all other uses, in any current or future media, including reprinting/republishing this material for advertising or promotional purposes, creating new collective works, for resale or redistribution to servers or lists, or reuse of any copyrighted component of this work in other works.}
}

\maketitle

\begin{abstract} 
The problem of measuring the best linear approximation of a nonlinear system by means of multilevel excitation sequences is analyzed. 
A comparison between different types of sequences applied at the input of Wiener systems is provided by numerical simulations and by experiments on a practical circuit including an analog filter and a clipping nonlinearity. 
The performance of the sequences is compared with a white Gaussian noise signal for reference purposes. 
The theoretical characterization of the best linear approximation when using randomized constrained sequences is derived analytically for the cubic nonlinearity case.
Numerical and experimental results show that the randomized constrained approach for designing ternary sequences has a low sensitivity to both even and odd order nonlinearities, resulting in a response close to the actual response of the underlying linear system.
\end{abstract}



%
\IEEEpeerreviewmaketitle


\section{Introduction}

The measurement of the frequency response function (FRF) of a linear dynamical system is a fundamental step that is typically performed in engineering applications for modeling or control purposes. 
Such procedure is typically performed by exciting the system under test with a properly designed input signal and measuring the output. 
However, many practical systems are affected by some degree of nonlinear distortion. For such systems, the FRF will vary as a function of the excitation that the system is subject to. To analyze these cases, linearization is often a viable option. For this reason, the concept of best linear approximation (BLA) of nonlinear systems is of great importance to practical applications. Therefore, the BLA, which is obtained by solving a least squares problem where the mean squared difference between the actual output of the system and the output of a linear model is minimized, has been thoroughly studied in the literature \cite{Evans&Rees2000_1, Ljung2001, Makila2006, EsfahaniEtAl2016}.

An analysis of the BLA for several cases of the amplitude distribution of the input signal is developed in \cite{WongEtAl2012}, considering Gaussian signals and binary signals. The analysis is further extended in \cite{WongEtAl2013} to multilevel signals. 
A particularly useful class of multilevel signals is that of ternary sequences, which are easy to generate in practical applications where the number of available levels is limited, and allow for defining a wider range of harmonic behavior when compared to binary sequences. 
In this context, a direct synthesis (DS) method  for obtaining ternary sequences with harmonic multiples of two and three suppressed is proposed in \cite{Tan2013}. Such sequences are well suited for the characterization of even and odd nonlinearities in dynamical systems and for the mitigation of the impact of nonlinear distortion on the FRF measurement. Furthermore, in \cite{DeAngelisEtAl2016, DeAngelisEtAl2016_TIM}, randomized constrained sequences (RCS) are introduced, allowing for a more flexible approach to design ternary sequences with defined spectral properties. The RCS approach is particularly suited for practical measurement applications, since it provides robustness in the presence of non-idealities such as nonuniform digital-to-analog converter (DAC) levels. 

In this paper, the study of the properties of RCS and DS ternary sequences is developed by analyzing their behavior when used as excitation signals to measure the BLA of a nonlinear system. The performance is evaluated on a block-structured nonlinear system and compared to commonly-used binary sequences and Gaussian noise. The latter is assumed as a reference, due to its well-studied properties for BLA measurement \cite{Enqvist&Ljung2005}.

The BLA is characterized theoretically, extending the numerical simulation results in \cite{DeAngelisEtAl2017I2MTC}.
Specifically, the analytical expression of the cross-correlation between the input and the output of Wiener systems with cubic nonlinearity and RCS input is derived. 
Furthermore, the behavior of the BLA is validated by experimental measurements performed on the prototype of a Wiener system implemented by an analog filter followed by a diode-based clipper.
It is shown that the RCS approach is relatively more robust to nonlinearities compared to the other considered approaches, thus providing a BLA closer to the underlying linear system.

\section{The BLA of a Wiener Nonlinear System}

The BLA of a nonlinear dynamical system is defined as a linear system such that its impulse response minimizes the mean square value of an error sequence $e\left[\cdot\right]$ defined as \cite{Pintelon&Schoukens2012}:
\begin{align}
e\left[k\right] = y\left[k\right] - \left(g_{BLA} \ast u\right)\left[k\right]
\end{align}
where $g_{BLA}\left[\cdot\right]$  is the impulse response of the BLA, $u\left[\cdot\right]$ is the input of the nonlinear system, $y\left[k\right]$ is its output at time instant $k$, and $\ast$ denotes the convolution operation.

\begin{figure}
\centering
\includegraphics[width = 0.8 \columnwidth]{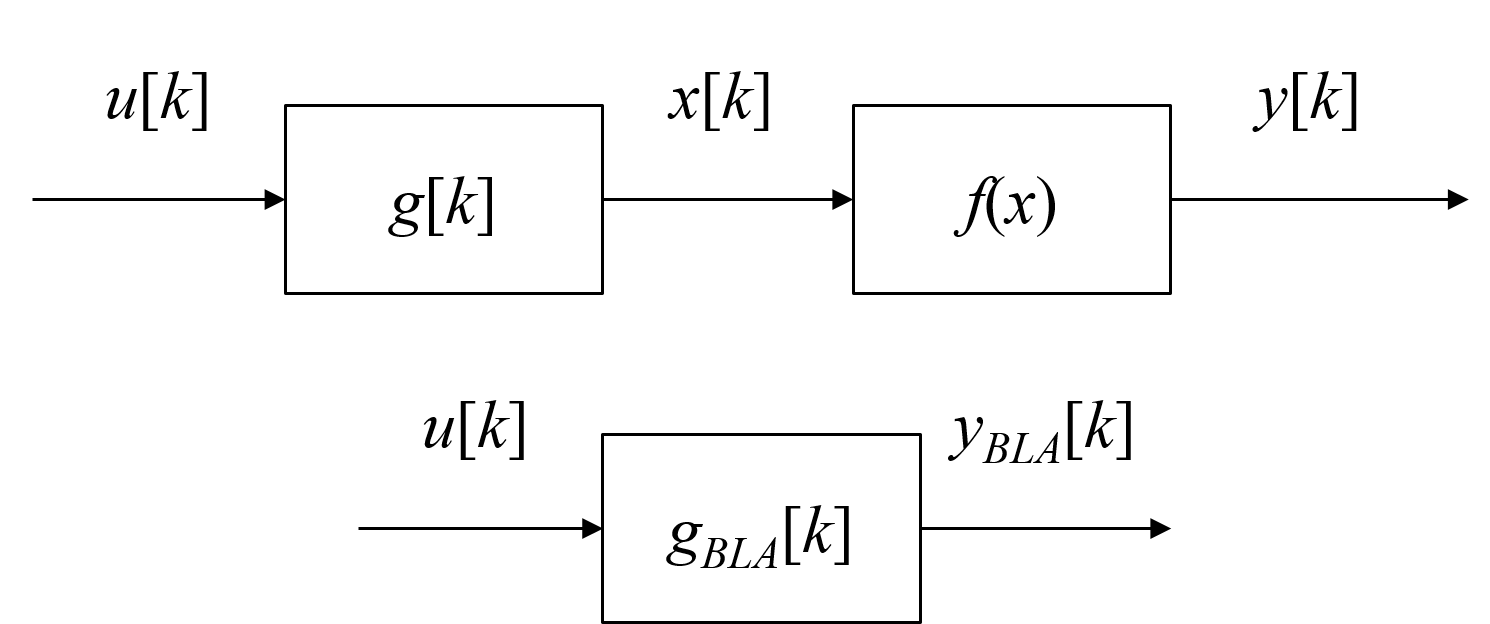}
\caption{Diagram of a Wiener system (top) and its BLA (bottom).
The symbol $g[k]$ denotes the impulse response of a linear system and the symbol $f(x)$ denotes a static nonlinear transformation acting on the sequence $x$.
\label{fig:diagram}}
\end{figure}

When the system under test is subject to different excitation signals, a different BLA is obtained, depending on the distribution of the amplitude of the excitation and its power spectrum. 
For a Wiener system, defined as the cascade of a linear dynamical system and a static nonlinearity as shown in Fig.~\ref{fig:diagram}, Gaussian excitations result in a BLA that is a scaled version of the underlying linear system \cite{WongEtAl2012}. 

This property represents a significant advantage in measurement applications, especially those where the dynamics of the underlying linear system must be measured while mitigating the impact of nonlinear distortions \cite{Evans&Rees2000_2}. However, the usage of Gaussian signals is not always advisable in practical scenarios. 
In fact, other classes of excitations are characterized by useful properties, such as a lower crest factor, resulting in more power into the system under test for the same maximum amplitude. 
Furthermore, in some applications, there is a limitation on the number of levels in the input signal. 
Such constraints may be related to the particular measurement process, ease of implementation considerations, or design of the actuator \cite{Barker&Godfrey1999}. 
Therefore, in those applications, one may resort to the use of binary sequences or multilevel signals.
The latter are defined as discrete sequences with a number of amplitude levels greater than two \cite{WongEtAl2013}. 
Ternary signals, which use three amplitude levels, are an example of a multilevel signal.

The use of signals with a limited number of levels implies that the distribution of the input deviates from Gaussianity. This causes the BLA to be biased with respect to the Gaussian BLA. This, in turn, means that the BLA will not be a scaled version of the linear system response. An in-depth theoretical study of such modifications is presented in \cite{WongEtAl2012}, which provides analytical expressions for the BLA using common non-Gaussian signals for general Wiener--Hammerstein systems.

\section{Excitation signals for measuring the BLA}
In this section, we provide a description of several signals that will then be used in the next section for numerical comparison of the BLA of a Wiener system.
The aim is to prove the applicability of such signals to the purpose of measuring the BLA, and to highlight the advantages of ternary sequences with respect to commonly-used binary sequences.

The excitation signals considered in this paper are ternary sequences, specifically DS and RCS, pseudorandom binary sequences, specifically maximum length binary sequences (MLBS) and inverse-repeat MLBS (IRMLBS), together with white Gaussian noise (WGN).

In particular, DS sequences are constructed according to the analytical method introduced in \cite{Tan2013}. 
A DS sequence is obtained starting from a basic sequence with specified harmonic properties, which can be a MLBS among other classes of sequences.
The basic sequence is repeated $n$ times and multiplied by a special fixed sequence. For suppression of harmonic multiples of 2 and 3, $n = 6$.
Due to this synthesis mechanism, the length $N$ of a MLBS-based DS must satisfy the constraint $N = 6(2^{2k+1}-1)$, with $k$ integer.
This requirement implies a lack of flexibility in terms of available periods. 
On the other hand, the advantage of the DS approach is that harmonic
multiples of two and three are suppressed, which removes the effects of even order nonlinearities and mitigates those of odd
order ones \cite{Tan2013, SchoukensEtAl2016}.

Recently, another approach, i.e. Randomized Constrained Sequences (RCS), for generating ternary sequences with harmonic multiples of two and three suppressed has been proposed in \cite{DeAngelisEtAl2016, DeAngelisEtAl2016_TIM}. Such approach is based on random sequences that are constrained to satisfy the harmonic suppression conditions and are iteratively modified in order to approach a design goal, such as uniformity of the levels of excited frequencies. The RCS method allows for a greater flexibility in terms of achievable sequence lengths, since it can provide sequences of length $6k$, with $k$ integer.

For comparison purposes, MLBS are considered in this paper, since these binary pseudorandom sequences are commonly used for measuring the response of dynamical systems. These sequences are generated using linear-feedback shift registers with properly chosen feedback terms \cite{GodfreyEtAl2005}. As a consequence, the available MLBS sequence lengths are $2^{k}-1$, with $k$ integer. Furthermore, there is a limited number of unique sequences for a given length. Notice that it is not possible to suppress odd-order harmonics using MLBS (given that they have only two levels). All harmonics are excited when using an MLBS, but even harmonic suppression can be achieved by IRMLBS, obtained by inverting every other digit in an MLBS of period $N$ to obtain an IRMLBS of period $2N$ \cite{TanEtAl2005}. For this reason, IRMLBS are also added to the comparison. 


\begin{figure*}
\centering
\    \ref{legend_location1} 
\vskip 0.1cm
\subfigure{
\begin{tikzpicture}
\input{plot_cubic_G_system1.tex}	
\end{tikzpicture}
}
\subfigure{
\begin{tikzpicture}
\input{plot_cubic_G_system2.tex}	
\end{tikzpicture}
}
\\
\vskip -0.1 cm
\subfigure{ \hskip 2 mm
\begin{tikzpicture}
\input{plot_cubic_G_system1_std.tex}	
\end{tikzpicture}
}
\subfigure{ \hskip 3 mm
\begin{tikzpicture}
\input{plot_cubic_G_system2_std.tex}	
\end{tikzpicture}
}
\\
\vskip -0.2 cm
\subfigure{
\begin{tikzpicture}
\input{plot_cubic_G_system3.tex}	
\end{tikzpicture}
}
\subfigure{
\begin{tikzpicture}
\input{plot_cubic_G_system4.tex}	
\end{tikzpicture}
}
\\
\vskip -0.1 cm
\subfigure{\hskip 3 mm
\begin{tikzpicture}
\input{plot_cubic_G_system3_std.tex}	
\end{tikzpicture}
}
\subfigure{\hskip 4 mm
\begin{tikzpicture}
\input{plot_cubic_G_system4_std.tex}	
\end{tikzpicture}
}
\vskip -0.2 cm
\caption{Numerical simulation results. The absolute value and the standard deviation of the estimated BLA are shown for the systems defined in Table I: (a) System 1; (b) System 2; (c) System 1 standard deviation; (d) System 2 standard deviation; (e) System 3; (f) System 4; (g) System 3 standard deviation; (h) System 4 standard deviation.
\label{fig:results}}
\end{figure*}
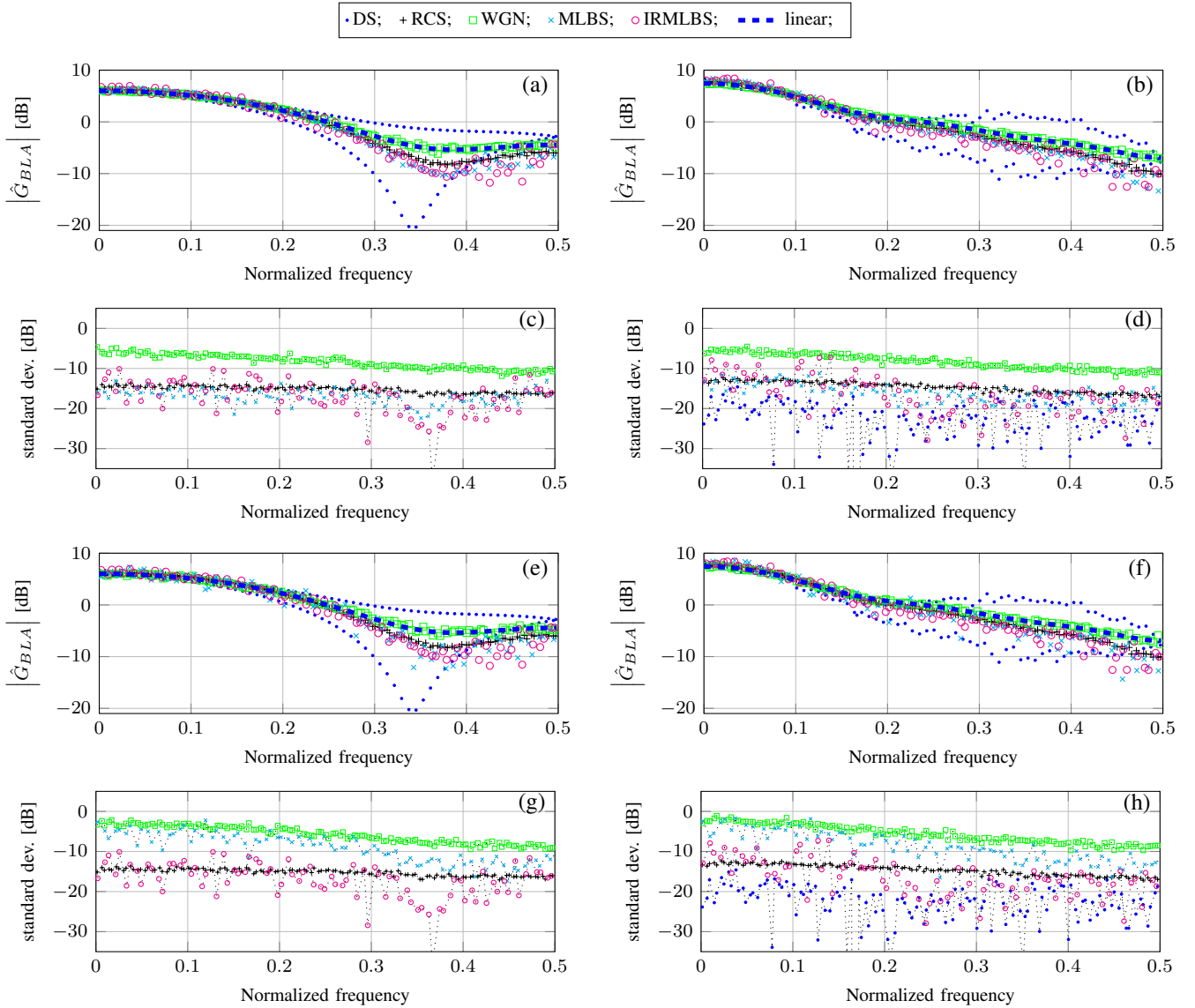

\section{Numerical comparison}
\label{sec:simulations}

For numerical simulation, four Wiener systems are considered, with the parameters shown in Table I. 
In particular, the linear block of the Wiener system consists of a FIR filter having impulse response coefficients given by [1 0.7 0.3] in the case of memory length 3, and [1 0.7 0.3 0.2 0.1 0.05] in the case of memory length 6. Furthermore, for the nonlinear block of the system, two nonlinearities are considered: $x^3$ and $x^3+x^2$.
For each system, each of the considered sequences is applied at the input, $u[k]$, and the output sequence $y[k]$ is computed at steady state, by discarding the transient. 
The system is noise free, because we are interested in studying the effect of the sequence distribution and not of measurement noise.
The sequence length is 762 for DS, RCS, and WGN, 511 for the MLBS case, and 510 for the case of IRMLBS.
Different lengths are used because of the sequence-specific length limitations explained in the previous section.
As an example, the length of DS must be $6(2^{2k+1}-1)$, which is an even number. 
On the other hand, the MLBS must be of the form $2^{k}-1$, which is an odd number. 
Therefore, it is not possible to generate a DS sequence and an MLBS sequence having the same length.
The sequence lengths were selected to be as close as possible. 
First, a value of 762 was chosen arbitrarily for the sequence length of DS. This choice allowed the computational complexity of further processing operations to be contained.
Once a length of 762 was defined for DS, the closest MLBS and IRMLBS lengths were chosen, namely 511 for MLBS and 2$\times$255 = 510 for IRMLBS.

{
\renewcommand{\arraystretch}{1.5}
\begin{table}
\centering
\caption{Summary of considered systems.}
\begin{tabular}{ | c | r | r |}
\hline
    & Impulse response $g[\cdot]$ & Nonlinearity $f(\cdot)$ \\ \hline
  System 1 & [1 0.7 0.3] & $x^3$ \\ \hline
  System 2 & [1 0.7 0.3 0.2 0.1 0.05] & $x^3$ \\ \hline
  System 3 & [1 0.7 0.3] & $x^3+x^2$ \\ \hline
  System 4 & [1 0.7 0.3 0.2 0.1 0.05] & $x^3+x^2$ \\ \hline
\end{tabular}
\end{table}
}

The BLA is estimated as follows \cite{Pintelon&Schoukens2012}:
\begin{align}
\hat{G}_{BLA} (j \omega_k) = \frac{\hat{S}_{YU}[k]}{\hat{S}_{UU}[k]}
\end{align}
where $\hat{S}_{YU}[k]$ is the cross-power spectrum of the input and output sequences, and $\hat{S}_{UU}[k]$ is the autopower spectrum of the input. 
The estimates are obtained by first dividing the sequence realizations into groups of 4 unique sequences each. 
Then, each group is used to provide an estimate of the BLA by averaging before division.
Such averaging significantly reduces the variance increase due to the random behavior for Gaussian signals \cite{Pintelon&Schoukens2012}.
The mean and standard deviation of the BLA estimates provided by each group are then calculated.
Due to the limited availability of unique MLBSs for a given length, as described in \cite{NewWaveInstruments}, 4 groups are formed for the DS and IRMLBS cases, 12 groups for the MLBS case, and 100 groups for the other cases.

The input sequences are normalized so that they have the same RMS value. The normalization is performed such that the impulse response of the Gaussian BLA is equivalent to the true underlying linear impulse response, according to the procedure in \cite{WongEtAl2012}. Additionally, the ternary sequences RCS and DS are multiplied by the term $\sqrt{3/2}$, in order to obtain the same power for all excitations.

Numerical results for all considered systems and input sequence types are shown in Fig. \ref{fig:results}, where only excited harmonics are displayed. In the figure, scaled versions of the BLA are plotted. The scaling factor is chosen such that the least square errors between each BLA and the FRF of the linear filter are minimized. This choice has been made because we are interested in the dissimilarity between the dynamics. The standard deviation has been scaled by the same factor.

\begin{figure*}
\centering
\    \ref{legend_location2} 
\vskip 0.1cm
\subfigure{
\begin{tikzpicture}
\input{plot_cubic_G_ratio_system1.tex}	
\end{tikzpicture}
}
\subfigure{
\begin{tikzpicture}
\input{plot_cubic_G_ratio_system2.tex}	
\end{tikzpicture}
}
\\
\vskip -0.1 cm
\subfigure{
\begin{tikzpicture}
\input{plot_cubic_G_ratio_system3.tex}	
\end{tikzpicture}
}
\subfigure{
\begin{tikzpicture}
\input{plot_cubic_G_ratio_system4.tex}	
\end{tikzpicture}
}
\vskip -0.2 cm
\caption{Numerical simulation results. Ratio between linear system response and BLA.  (a) System 1; (b) System 2; (c) System 3; (d) System 4.\label{fig:results_ratio}}
\end{figure*}
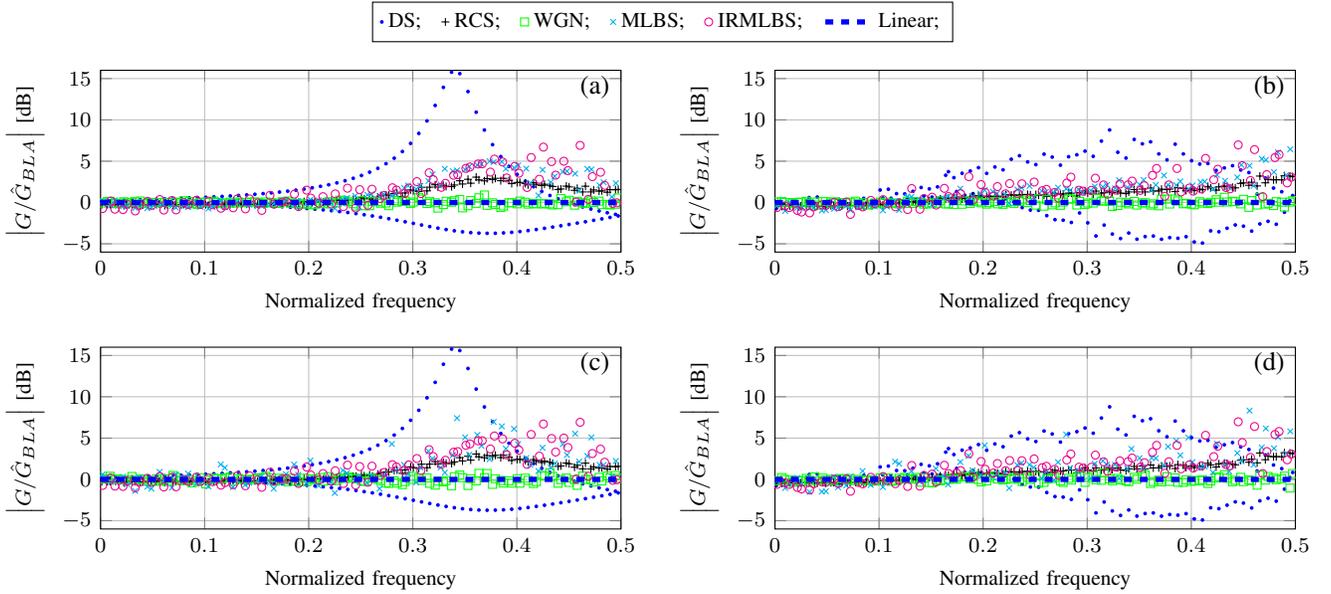

In Fig. \ref{fig:results_ratio}, the ratio between the linear system response and the BLA is plotted, with the same scaling as in Fig. \ref{fig:results}, from which it can be seen that the RCS input signal provides the closest response to the true underlying linear system response, except for Gaussian noise.
This is because the amplitude distribution of the RCS at the output of the linear system is the closest to a Gaussian distribution, among the considered signals. The latter behavior is also highlighted by the 
Kullback-Leibler divergence \cite{Kullback&Leibler1951} values listed in Table~\ref{tab:KLD}.

\begin{table}[]
\renewcommand{\arraystretch}{1.3}
\caption{Kullback-Leibler divergence between the empirical probability distributions, estimated by histograms, of the considered sequences and the ideal normal distribution, at the output of the linear block of System 2.}
\label{tab:KLD}
\centering
\begin{tabular}{|l|l|l|l|l|}
\hline
DS & RCS & WGN & MLBS & IRMLBS \\
\hline
0.2346 &   0.0654 &   0.0255 &   0.1837 &   0.1897\\
\hline
\end{tabular}
\end{table}


\section{Discussion of the results}

Results show that odd ternary sequences are less sensitive to the effect of even-order nonlinearities and memory length of the linear system. In fact, they show less variability when compared to white Gaussian noise excitation and to binary signals that excite all harmonics (such as MLBS). This can be noticed by observing that the standard deviation for MLBS in Fig. \ref{fig:results}(g), where the $x^2$ term is present, is considerably larger than that of Fig. \ref{fig:results}(c).

Moreover, from the results of Fig.~\ref{fig:results_ratio}, it can be noticed that DS is very sensitive to nonlinear distortions, due to its particular construction. On the other hand, RCS, thanks to its randomization properties, is less sensitive to nonlinearity. Still, RCS maintains the ability to discriminate even and odd distortions, due to harmonic multiples of two and three suppressed. This is not possible with MLBS and IRMLBS.

\section{Theory for cubic nonlinearity and RCS input}
\label{sec:theory}
Motivated by the numerical simulation results presented in the previous sections, we analyze the performance of RCS for Wiener systems also from a theoretical point of view.
Specifically,
in this section, we provide a theoretical characterization of the behavior of RCS when applied to the input of a Wiener system with a cubic nonlinearity. 
Notice that a cubic nonlinearity may be used to model numerous real-world nonlinear systems often encountered in practice, which can be represented using odd functions.
The theoretical characterization is performed by deriving the cross-correlation between the input and the output, as stated in the following Theorem. 

\begin{theorem}
Consider an RCS  of length $N$ as defined in \cite{DeAngelisEtAl2016_TIM} at the input of a causal Wiener system. Let the Wiener system consist of a FIR filter of order $H$, with $H<N/6$, and a cubic nonlinearity. Denote the input of the Wiener system by $u[\cdot]$, its output by $y[\cdot]$, the autocorrelation of the input by $R_u[\cdot]$, and the impulse response of the FIR filter by $g[\cdot]$.

Then, the cross-correlation between the output and the input is given by
\begin{align}
R_{yu}[r] & = 2 \alpha_2 \left( g \ast R_u \right) [r] - \left( g^3 \ast R_u \right) [r] \label{eq:theorem1}\\
& = g\left[r_m\right] R_u\left[ r_q \right] 
\left( 2\alpha_2 - g^2\left[r_m\right] \right)\label{eq:theorem2} \,,\\
r & = 0 \ldots N-1 \,, \notag
\end{align}
\noindent 
where $\alpha_2 \triangleq \sum\limits_{k=0}^H g^2\left[k\right]$,
$r_q$ denotes the quotient of the division $\frac{r}{N/6}$, i.e. 
$r_q \triangleq \left\lfloor \frac{r}{N/6} \right\rfloor \frac{N}{6}$, 
and $r_m$ denotes the reminder, with
$r_m = r - r_q$.
%
\end{theorem}

\begin{IEEEproof}
See Appendix I.
\end{IEEEproof}

Note that the expressions for the cross-correlation in \eqref{eq:theorem1} and \eqref{eq:theorem2} apply also to Wiener systems where the nonlinearity is $y=x^3+x^2$.
This is due to the general result that zero-mean excitations with symmetric distribution, combined with even-order nonlinearities (with even-degree Volterra kernels), have a zero BLA \cite{Pintelon&Schoukens2012}.

It is interesting to compare the derived expression of the cross-correlation to that for white Gaussian input sequences of variance one, which is given by \cite{WongEtAl2012}:
\begin{align}
\label{eq:R_Gaussian}
R_{yu(Gaussian)}[r] = 3 \alpha_2 g[r] \,.
\end{align}
If the cross-correlation in \eqref{eq:theorem1} is evaluated for $\left| r
\right| < N/6$ and for an RCS input of unitary variance, i.e. the RCS sequence is scaled by $\sqrt{3/2}$, it reduces to 
\begin{align}
\label{eq:R_RCS}
R_{yu}[r] & = 3 \alpha_2 g[r] - \frac{3}{2} g^3 [r] .
\end{align}
By comparing \eqref{eq:R_RCS} to \eqref{eq:R_Gaussian}, it can be noticed that the bias term of the RCS with respect to the Gaussian input case is $- 3/2 g^3 [r]$, whereas for random binary sequences it is $-2 g^3 [r]$, which applies also to MLBS in the limit when $N$ is large \cite{WongEtAl2012}.
Therefore, the BLA of the RCS is closer to the underlying linear system than that of the MLBS.

\begin{figure}
\centering
\subfigure
   {\input{Fig_4.tex}}\\
   \hskip 1mm
\subfigure
   {\input{Fig_4_b.tex}}
\caption{
Comparison between numerical simulations and theoretical equations \eqref{eq:theorem1} and \eqref{eq:theorem2}. (a): full cross-correlation sequence. (b): magnification of the first peak of the cross-correlation sequence. 
\label{fig:comparison_theory_simulation}}
\end{figure}
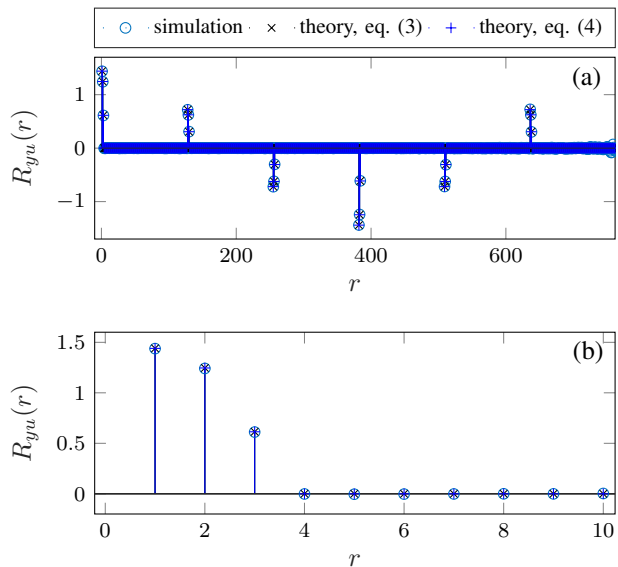

The theoretical results in Theorem 1 have been validated by numerical simulations. 
In particular, $10^3$ different instances of an RCS of length $N=762$ were generated and fed to a Wiener system with $H=3$, $g=\left[ 1 ,\, 0.7 ,\,  0.3 \right]$, and a cubic nonlinearity. 
For each input sequence, the corresponding output was stored. 
Subsequently, the cross-correlation between the input and the output was calculated for each sequence. 
Finally, the average of all cross-correlation sequences was calculated.
Such average cross-correlation is depicted in Fig. \ref{fig:comparison_theory_simulation}, together with the theoretical values obtained from \eqref{eq:theorem1} and \eqref{eq:theorem2}.
As shown in Fig. \ref{fig:comparison_theory_simulation}, there is a very good agreement between theoretic results and simulations, thus validating the derivations of Theorem 1. 
In the following section, we show that the behavior analyzed theoretically for cubic nonlinearity may be observed also in a practical situation, with a realistic type of nonlinearity.

\section{Experimental Results on a Real-World Wiener System}
\begin{figure}
\centering
\tikzstyle{every path}=[line width=0.8pt,line cap=round,line join=round]
\ctikzset{bipoles/length=1cm, bipoles/thickness=1}
\footnotesize
\begin{circuitikz}[european voltages, scale = 0.8, transform shape]
	\draw[color=black]
		(0,2) to[R=180 $\Omega$] (2,2) 
		to[D] (2,0)
		to (2.5,0) node[sground]{}
		(2,0) to (3,0)
		to[D] (3,2) 
		to (2,2)
		(3,2) to[short,-o] (3.5,2) 
		(3,0) to[short,-o] (3.5,0) 
		(3.5,2) to[open, v^=$V_{O}$] (3.5,0)
		
		(-1,2) node[op amp] (opamp) {}
		(opamp.+) node[left] {}
		(opamp.-) node[left] {}
		(opamp.out) node[right] {}
		(opamp.+) to node[sground]{} (-2.2,1)
			
		(0,2) to[short, *-]
		(0,3.2) to[R, l_=1 k$\Omega$]
		(-2.2,3.2) to [short, -*]
		(-2.2,2.5) to [short] (opamp.-)
		
		(-2.2,3.2) to [short]
		(-2.2,4.2) to [C, l=100 nF]
		(0,4.2) to [short]
		(0,3.2)
		
		(-2.2,2.5) to [R=1 k$\Omega$, -o] (-4,2.5)
		(-4,1) to [short, o-] (-4,1)
		(-4,1) to node[sground]{} (-4,0.5)
		
		(-4,2.5) to[open, v_=$V_{I}$] (-4,1)
;
\end{circuitikz}
\caption{Schematic of the realized Wiener system, consisting of a low pass filter and a soft clipper.\label{fig:schematic}}
\end{figure}
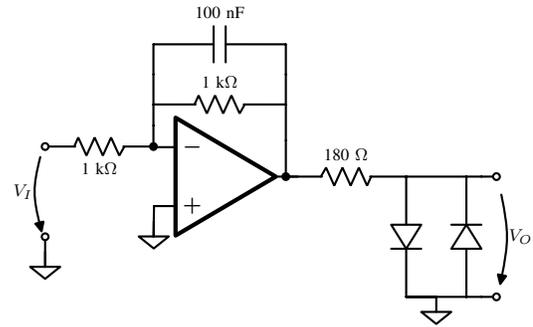

To validate the theoretical results presented in Sections~\ref{sec:simulations}~-~\ref{sec:theory}, experiments were performed on a real-world Wiener system, implemented as an electronic circuit.
The considered Wiener system is comprised of a first-order low-pass active filter, having a cutoff frequency of approximately 1600 Hz, followed by a diode-based clipper with a series resistor, thus performing soft clipping, as depicted in the schematic of Fig.~\ref{fig:schematic}.
This circuit was first analyzed by numerical simulations, then experimentally tested. 

\subsection{Simulation}
\begin{figure*}[t]
\centering
\    \ref{legend_location3}
\vskip 0.1cm
\subfigure{
\begin{tikzpicture}
\input{plot_clip_G_ratio_A1.tex}	
\end{tikzpicture}
}
\subfigure{
\begin{tikzpicture}
\input{plot_clip_G_ratio_A2.tex}	
\end{tikzpicture}
}
\\
\vskip -0.25cm
\subfigure{
\begin{tikzpicture}
\input{plot_clip_G_phase_diff_A1.tex}
\end{tikzpicture}
}
\subfigure{
\begin{tikzpicture}
\input{plot_clip_G_phase_diff_A2.tex}
\end{tikzpicture}
}
\caption{
Wiener system of Fig.~\ref{fig:schematic}, numerical simulation results.
The two graphs on the left are obtained with an input RMS level of 1 V, whereas those on the right with an input RMS level of 2 V, thus providing a stronger excitation of the nonlinearity. 
\label{fig:clipping_simulation}
}
\end{figure*}
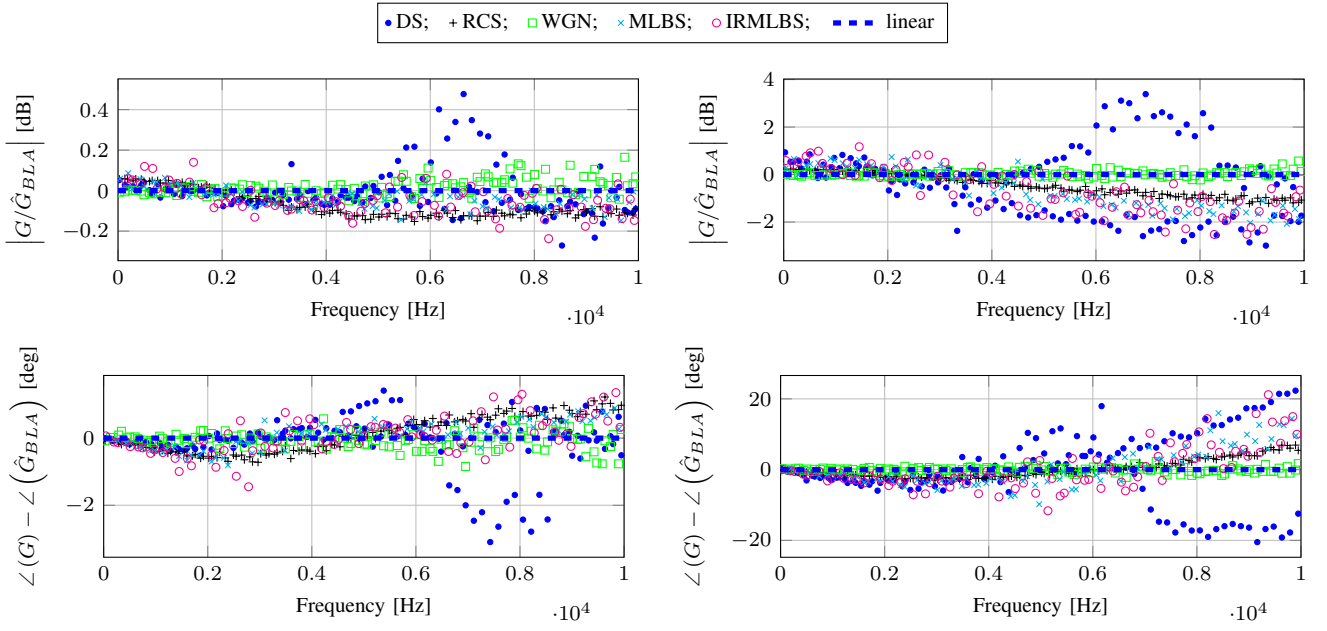

Numerical simulations were carried out by modeling the lowpass filter as a first-order iir filter and performing hard clipping numerically. Hard clipping is an approximation of the soft clipping operation of the actual circuit.
The same simulation procedure as in Section~\ref{sec:simulations} was used. Specifically, due to a limited number of sequences available, 16 instances of DS and IRMLBS, 48 instances of MLBS, and 1000 instances of RCS and WGN were used.
Oversampling by a factor of 10 was employed, to mitigate the effect of frequency warping close to Nyquist frequency due to the iir filter. 
Therefore, the sampling frequency of the data acquisition system was 200~kSa/s, whereas the signal generator frequency was 20~kHz.
Simulation results are shown in Fig. \ref{fig:clipping_simulation}, where plots are scaled so that the mean squared error between each BLA and the FRF of the linear filter is minimized.

\subsection{Experimental Characterization}
A prototype of the considered Wiener system shown in Fig.~\ref{fig:schematic} was built using two 1N4148 diodes and a OP27 operational amplifier. Experiments were conducted using the ``Analog discovery 2'' acquisition board by Digilent \cite{Digilent}. 
In particular, the same sequences as those used in simulations were applied at the input, scaled so they had the same desired RMS amplitude, and the input and output of the system were measured simultaneously, discarding the initial transient.
The experiment was repeated for several values of the RMS amplitude.
The sampling frequency of the on-board DAC was set to 20 kSa/s, with a resolution of 14 bits. 
Moreover, the sampling frequency of the ADC was set to 200 kSa/s with a resolution of 14 bits. 
Note that oversampling was performed also in the case of the experiments. 
The purpose of oversampling in this context was to mitigate the effect of sharp transitions of the input signal.

As a reference, the frequency response of the underlying linear system was also measured, using a stepped sine excitation with a low amplitude of 0.1 V that was chosen in order to avoid exciting the nonlinearity.

\begin{figure*}[t]
\centering
\    \ref{legend_location4}
\vskip 0.1cm
\subfigure{
\begin{tikzpicture}
\input{plot_exper_clip_G_ratio_A1.tex}	
\end{tikzpicture}
}
\subfigure{
\begin{tikzpicture}
\input{plot_exper_clip_G_ratio_A2.tex}	
\end{tikzpicture}
}
\\
\vskip -0.25cm
\subfigure{
\begin{tikzpicture}
\input{plot_exper_clip_G_phase_diff_A1.tex}
\end{tikzpicture}
}
\subfigure{
\begin{tikzpicture}
\input{plot_exper_clip_G_phase_diff_A2.tex}
\end{tikzpicture}
}
\caption{
Wiener system of Fig.~\ref{fig:schematic}, experimental test results for an input RMS level of 1 V (left) and 2 V (right).
\label{fig:clipping_experiments}
}
\end{figure*}
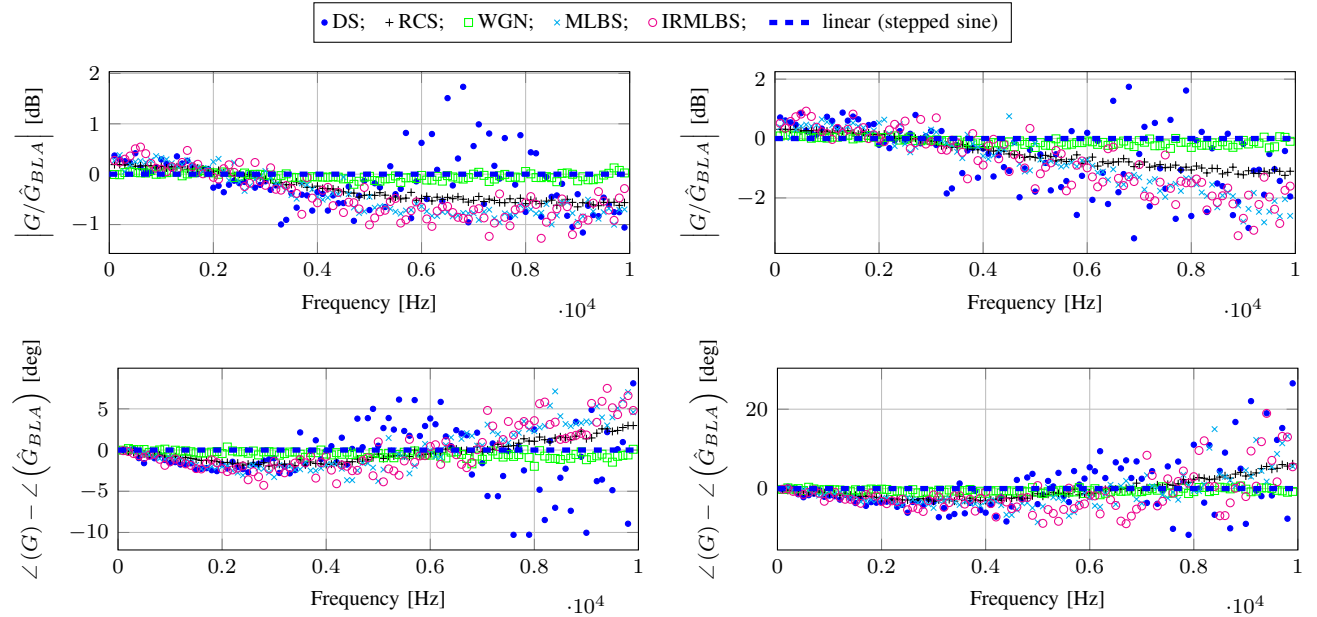

Experimental results are shown in Fig.~\ref{fig:clipping_experiments}, where the averaging process and number of sequence instances used is the same as in the simulations.
By comparing such results with numerical simulations in Fig.~\ref{fig:clipping_simulation}, a strong agreement in the behavior of the considered sequences can be noticed.
Specifically, the RCS is closer to the underlying linear system, both in terms of magnitude and phase, with respect to the other multilevel sequences compared. 
This is particularly emphasized when the nonlinearity is strongly excited.

\section{Conclusion}
The use of ternary sequences for measuring the BLA of a Wiener nonlinear system was investigated by numerical simulations, theoretical derivations, and experiments.
The theoretical characterization of the randomized constrained sequences was provided.
Moreover, the performance of ternary sequences was compared to that of commonly used binary sequences and to white Gaussian noise. 
The comparison was carried out both numerically and by experiments on a realized Wiener circuit performing analog filtering and clipping.
Results show that the randomized constrained sequence approach provides low sensitivity to nonlinearities,
resulting in a BLA that is close to the underlying linear system response.
Compared to binary sequences, it has the advantage of suppressing harmonic multiples of two and three, allowing a nonparametric nonlinear analysis to be made without any user interaction.
Compared to other signals that provide such suppression, such as multisines, RCS is also robust to DAC nonlinearities, while a multisine requires highly linear DAC for proper operation.

\appendices
\section{Proof of Theorem 1}
The cross-correlation between the output and the input is
\begin{align}
R_{yu}[r] & = E\left[ y[t] u[t-r] \right] \notag \\
& = E\left[ \left( \sum_{k=0}^{H-1} g[k] u[t-k] \right)^3 u[t-r] \right] \label{eq:xcorr_initial}
\end{align}

By expanding \eqref{eq:xcorr_initial}, we obtain
\begin{align}
& R_{yu}[r] = 
	E\left[ 
	\left( \vphantom{\sum_{j = h+1}^{H-1} }
		\sum_{k=0}^{H-1} g^3[k] u^3[t-k] \notag \right. \right.\\
		& \left. \left. 
		\,+ 3 \sum_{k=0}^{H-1} \sum_{h = k+1}^{H-1} g^2[k] g[h] u^2[t-k] u[t-h] 
		\right. \right. \notag \\
		& \left. \left. 
		\,+ 6 \sum_{k=0}^{H-1}
		\sum_{h=k+1}^{H-1}
		\sum_{j=h+1}^{H-1} 
		g[k] g[h] g[j] u[t-k] u[t-h] u[t-j] 
	\right) \right. \notag \\
	& \left. \,\times u[t-r] 
	\vphantom{\sum_{j = h+1}^{H-1} }
	\right] \,, \quad 	r = 0 \ldots N-1
	\label{eq:xcorr1}
\end{align}
where the terms of the expansion of a polynomial raised to the third power are given by the multinomial coefficients \cite{DLMF}. 

Recall from \cite{DeAngelisEtAl2016_TIM} that $E\left[ u[n] \right] = 0 \,, \forall n$ and that the autocorrelation of the RCS is
\begin{align}
\label{eq:autocorrelation_ideal}
R_u[n] = \left\{ 
\begin{array}{ll}
\frac{2}{3} & n=0 \\
-\frac{2}{3} & n=\frac{N}{2} \\
\frac{1}{3} & n=\frac{N}{6} \,, \frac{5}{6} N\\
-\frac{1}{3} & n=\frac{N}{3} \,, \frac{2}{3} N\\
0 & \mbox{elsewhere}
\end{array}
\right.  \,.
\end{align}
By using \eqref{eq:autocorrelation_ideal}, and considering that $H<N/6$, we obtain, for $k,h,j=0 \ldots H-1 , \, h \neq k , \, j \neq k , \, j \neq h $, that the random variables $u\left[t-k\right]$, $u[t-h]$, $u[t-j]$, and $u[t-r]$ are statistically independent and 
\begin{align}
\begin{split}
E\left[  u\left[t-k\right] u[t-h] u[t-j] u[t-r] \right] = 0 \,.
\end{split}
\label{eq:E_zero}
\end{align}

Then, by substituting \eqref{eq:E_zero} in \eqref{eq:xcorr1} and noting that $u^3[\cdot] \equiv u[\cdot]$ for any ternary sequence assuming values $[1\,, -1 \,, 0]$, the expression in \eqref{eq:xcorr1} for the cross-correlation simplifies to:
\begin{align}
& R_{yu}[r] =  E\left[ 
		\sum_{k=0}^{H-1} g^3[k] u[t-k] u[t-r] \notag \right. \\
		& \left. 
		\,+ 3 \sum_{k=0}^{H-1} \sum_{\substack{h=0 \\ h \neq k}}^{H-1} 
		g^2[k] g[h] u^2[t-k] u[t-h] u[t-r]
		\right] \notag \\
& = \sum_{k=0}^{H-1} g^3[k] R_u[r-k] \notag \\
		& \,+ 3 \sum_{k=0}^{H-1} \sum_{\substack{h=0 \\ h \neq k}}^{H-1} 
		g^2[k] g[h] E\left[u^2[t-k]\right] R_u[r-h] \notag \\
& = \sum_{k=0}^{H-1} g^3[k] R_u[r-k]
	+ 2 \sum_{k=0}^{H-1} \sum_{\substack{h=0 \\ h \neq k}}^{H-1} 
		g^2[k] g[h] R_u[r-h]
		\label{eq:xcorr2}
\end{align}
where we used the fact that the variance of the RCS is $E\left[u^2[n]\right] = 2/3 \,, \forall n$, and the result that functions of independent random variables are independent.

By using simple algebraic manipulations, \eqref{eq:xcorr2} can be written in terms of the convolution operator as follows:
\begin{align}
R_{yu}[r] & = \left( g^3 \ast R_u \right) [r] 
+ 2 \sum_{k=0}^{H-1} \sum_{\substack{h=0 \\ h \neq k}}^{H-1} g^2[k] g[h] R_u[r-h] \notag \\
& =  \left( g^3 \ast R_u \right) [r] 
+ 2 \left(\sum_{k=0}^{H-1} g^2[k]\right) \sum_{h=0}^{H-1} g[h] R_u[r-h] \notag \\
& \quad - 2 \sum_{k=0}^{H-1} g^2[k] g[k] R_u[r-k] \notag \\
& =  \left( g^3 \ast R_u \right) [r]
 + 2 \left(\sum_{k=0}^{H-1} g^2[k]\right) \left( g \ast R_u \right) [r]  \notag \\
& \quad - 2 \left( g^3 \ast R_u \right) [r] \notag \\
& = 2 \alpha_2 \left( g \ast R_u \right) [r] - \left( g^3 \ast R_u \right) [r]
\notag 
\end{align}

Therefore, \eqref{eq:theorem1} is proven. 
Finally, the expression in \eqref{eq:theorem2} follows from \eqref{eq:theorem1} by noticing that only one term of the convolution sums in \eqref{eq:theorem1} is nonzero for a given $r$.


\end{document}

%% file: plot_cubic_G_system1.tex
	
\begin{axis}[
grid=major, 
xmin=0, xmax=0.5,	
ymin=-21, ymax=10,
width=8.5cm, height=4.00 cm, 
legend columns=-1,
legend to name=legend_location1,
xlabel={\footnotesize Normalized frequency},
ylabel={\footnotesize $\left|\hat{G}_{BLA}\right|$ [dB]},
clip marker paths=true
]	
	
	\node at (0.475,7) {(a)};

	\addplot 
	[mark=*, mark size=0.5pt, only marks, 
		mark options={draw=blue,fill=blue, solid}] 
	table[x expr=\thisrowno{0}, y expr=\thisrowno{1}, col sep =comma] 
	{./data/data_cubic_system1_input1.txt};
	\addlegendentry{DS; \ \ }
	
	
	\addplot 
	[mark=+, mark size=1.5pt, only marks, 
		mark options={draw=black, fill=black, solid}] 
	table[x expr=\thisrowno{0}, y expr=\thisrowno{1}, col sep =comma] 
	{./data/data_cubic_system1_input2.txt};
	\addlegendentry{RCS;  \ \ }	
	
	
	\addplot 
	[mark=square, mark size=1.5pt, only marks, 
		mark options={draw=green, fill=green, solid}] 
	table[x expr=\thisrowno{0}, y expr=\thisrowno{1}, col sep =comma] 
	{./data/data_cubic_system1_input3.txt};
	\addlegendentry{WGN;  \ \ }	
	
	
	\addplot 
	[mark=x, mark size=1.5pt, only marks, 
		mark options={draw=cyan, fill=cyan, solid}] 
	table[x expr=\thisrowno{0}, y expr=\thisrowno{1}, col sep =comma] 
	{./data/data_cubic_system1_input4.txt};
	\addlegendentry{MLBS;  \ \ }	
	
	
	\addplot 
	[mark=o, mark size=1.5pt, only marks, 
		mark options={draw=magenta, fill=magenta, solid}] 
	table[x expr=\thisrowno{0}, y expr=\thisrowno{1}, col sep =comma] 
	{./data/data_cubic_system1_input5.txt};
	\addlegendentry{IRMLBS;  \ \ }
	
	
	\addplot 
	[dashed, draw=blue, line width=2pt] 
	table[x expr=\thisrowno{0}, y expr=\thisrowno{1}, col sep =comma] 
	{./data/data_cubic_system1_linear.txt};
	\addlegendentry{linear;\ \ }

\end{axis}
	

%% file: plot_cubic_G_system2.tex
	
\begin{axis}[
grid=major, 
xmin=0, xmax=0.5,	
ymin=-21, ymax=10,
width=8.5cm, height=4.00 cm, 
xlabel={\footnotesize Normalized frequency},
ylabel={\footnotesize $\left|\hat{G}_{BLA}\right|$ [dB]},
clip marker paths=true
]		
	\node at (0.475,7) {(b)};

	\addplot 
	[mark=*, mark size=0.5pt, only marks, 
		mark options={draw=blue,fill=blue, solid}] 
	table[x expr=\thisrowno{0}, y expr=\thisrowno{1}, col sep =comma] 
	{./data/data_cubic_system2_input1.txt};
	
	
	\addplot 
	[mark=+, mark size=1.5pt, only marks, 
		mark options={draw=black, fill=black, solid}] 
	table[x expr=\thisrowno{0}, y expr=\thisrowno{1}, col sep =comma] 
	{./data/data_cubic_system2_input2.txt};
	
	
	\addplot 
	[mark=square, mark size=1.5pt, only marks, 
		mark options={draw=green, fill=green, solid}] 
	table[x expr=\thisrowno{0}, y expr=\thisrowno{1}, col sep =comma] 
	{./data/data_cubic_system2_input3.txt};
	
	
	\addplot 
	[mark=x, mark size=1.5pt, only marks, 
		mark options={draw=cyan, fill=cyan, solid}] 
	table[x expr=\thisrowno{0}, y expr=\thisrowno{1}, col sep =comma] 
	{./data/data_cubic_system2_input4.txt};
	
	
	\addplot 
	[mark=o, mark size=1.5pt, only marks, 
		mark options={draw=magenta, fill=magenta, solid}] 
	table[x expr=\thisrowno{0}, y expr=\thisrowno{1}, col sep =comma] 
	{./data/data_cubic_system2_input5.txt};
	
	
	\addplot 
	[dashed, draw=blue, line width=2pt] 
	table[x expr=\thisrowno{0}, y expr=\thisrowno{1}, col sep =comma] 
	{./data/data_cubic_system2_linear.txt};

\end{axis}
	

%% file: plot_cubic_G_system1_std.tex
	
\begin{axis}[
grid=major, 
xmin=0, xmax=0.5,	
ymin=-35, ymax=5,
width=8.5cm, height=4.00 cm, 
legend columns=-1,
xlabel={\footnotesize Normalized frequency},
ylabel={\footnotesize standard dev. [dB]},
clip marker paths=true
]	
	
	\node at (0.475,2) {(c)};

	
	\addplot 
	[dotted, mark=*, mark size=0.5pt, 
		mark options={draw=blue,fill=blue, solid}] 
	table[x expr=\thisrowno{0}, y expr=\thisrowno{2}, col sep =comma] 
	{./data/data_cubic_system1_input1.txt};
	
	
	\addplot 
	[dotted, mark=+, mark size=1pt, 
		mark options={draw=black, fill=black, solid}] 
	table[x expr=\thisrowno{0}, y expr=\thisrowno{2}, col sep =comma] 
	{./data/data_cubic_system1_input2.txt};
	
	
	\addplot 
	[dotted, mark=square, mark size=1pt, 
		mark options={draw=green, fill=green, solid}] 
	table[x expr=\thisrowno{0}, y expr=\thisrowno{2}, col sep =comma] 
	{./data/data_cubic_system1_input3.txt};
	
	
	\addplot 
	[dotted, mark=x, mark size=1pt,
		mark options={draw=cyan, fill=cyan, solid}] 
	table[x expr=\thisrowno{0}, y expr=\thisrowno{2}, col sep =comma] 
	{./data/data_cubic_system1_input4.txt};
	
	
	\addplot 
	[dotted, mark=o, mark size=1pt,
		mark options={draw=magenta, fill=magenta, solid}] 
	table[x expr=\thisrowno{0}, y expr=\thisrowno{2}, col sep =comma] 
	{./data/data_cubic_system1_input5.txt};
	

\end{axis}
	

%% file: plot_cubic_G_system2_std.tex
	
\begin{axis}[
grid=major, 
xmin=0, xmax=0.5,	
ymin=-35, ymax=5,
width=8.5cm, height=4.00 cm, 
xlabel={\footnotesize Normalized frequency},
ylabel={\footnotesize standard dev. [dB]},
clip marker paths=true
]		
	\node at (0.475,2) {(d)};

	
	\addplot 
	[dotted, mark=*, mark size=0.5pt, 
		mark options={draw=blue,fill=blue, solid}] 
	table[x expr=\thisrowno{0}, y expr=\thisrowno{2}, col sep =comma] 
	{./data/data_cubic_system2_input1.txt};
	
	
	\addplot 
	[dotted, mark=+, mark size=1pt, 
		mark options={draw=black, fill=black, solid}] 
	table[x expr=\thisrowno{0}, y expr=\thisrowno{2}, col sep =comma] 
	{./data/data_cubic_system2_input2.txt};
	
	
	\addplot 
	[dotted, mark=square, mark size=1pt, 
		mark options={draw=green, fill=green, solid}] 
	table[x expr=\thisrowno{0}, y expr=\thisrowno{2}, col sep =comma] 
	{./data/data_cubic_system2_input3.txt};
	
	
	\addplot 
	[dotted, mark=x, mark size=1pt,
		mark options={draw=cyan, fill=cyan, solid}] 
	table[x expr=\thisrowno{0}, y expr=\thisrowno{2}, col sep =comma] 
	{./data/data_cubic_system2_input4.txt};
	
	
	\addplot 
	[dotted, mark=o, mark size=1pt,
		mark options={draw=magenta, fill=magenta, solid}] 
	table[x expr=\thisrowno{0}, y expr=\thisrowno{2}, col sep =comma] 
	{./data/data_cubic_system2_input5.txt};
	

\end{axis}
	

%% file: plot_cubic_G_system3.tex
	
\begin{axis}[
grid=major, 
xmin=0, xmax=0.5,	
ymin=-21, ymax=10,
width=8.5cm, height=4.00 cm, 
xlabel={\footnotesize Normalized frequency},
ylabel={\footnotesize $\left|\hat{G}_{BLA}\right|$ [dB]},
clip marker paths=true
]		
	\node at (0.475,7) {(e)};

	\addplot 
	[mark=*, mark size=0.5pt, only marks, 
		mark options={draw=blue,fill=blue, solid}] 
	table[x expr=\thisrowno{0}, y expr=\thisrowno{1}, col sep =comma] 
	{./data/data_cubic_system3_input1.txt};
	
	
	\addplot 
	[mark=+, mark size=1.5pt, only marks, 
		mark options={draw=black, fill=black, solid}] 
	table[x expr=\thisrowno{0}, y expr=\thisrowno{1}, col sep =comma] 
	{./data/data_cubic_system3_input2.txt};
	
	
	\addplot 
	[mark=square, mark size=1.5pt, only marks, 
		mark options={draw=green, fill=green, solid}] 
	table[x expr=\thisrowno{0}, y expr=\thisrowno{1}, col sep =comma] 
	{./data/data_cubic_system3_input3.txt};
	
	
	\addplot 
	[mark=x, mark size=1.5pt, only marks, 
		mark options={draw=cyan, fill=cyan, solid}] 
	table[x expr=\thisrowno{0}, y expr=\thisrowno{1}, col sep =comma] 
	{./data/data_cubic_system3_input4.txt};
	
	
	\addplot 
	[mark=o, mark size=1.5pt, only marks, 
		mark options={draw=magenta, fill=magenta, solid}] 
	table[x expr=\thisrowno{0}, y expr=\thisrowno{1}, col sep =comma] 
	{./data/data_cubic_system3_input5.txt};
	
	
	\addplot 
	[dashed, draw=blue, line width=2pt] 
	table[x expr=\thisrowno{0}, y expr=\thisrowno{1}, col sep =comma] 
	{./data/data_cubic_system3_linear.txt};

\end{axis}
	

%% file: plot_cubic_G_system4.tex
	
\begin{axis}[
grid=major, 
xmin=0, xmax=0.5,	
ymin=-21, ymax=10,
width=8.5cm, height=4.00 cm, 
xlabel={\footnotesize Normalized frequency},
ylabel={\footnotesize $\left|\hat{G}_{BLA}\right|$ [dB]},
clip marker paths=true
]		
	\node at (0.475,7) {(f)};

	\addplot 
	[mark=*, mark size=0.5pt, only marks, 
		mark options={draw=blue,fill=blue, solid}] 
	table[x expr=\thisrowno{0}, y expr=\thisrowno{1}, col sep =comma] 
	{./data/data_cubic_system4_input1.txt};
	
	
	\addplot 
	[mark=+, mark size=1.5pt, only marks, 
		mark options={draw=black, fill=black, solid}] 
	table[x expr=\thisrowno{0}, y expr=\thisrowno{1}, col sep =comma] 
	{./data/data_cubic_system4_input2.txt};
	
	
	\addplot 
	[mark=square, mark size=1.5pt, only marks, 
		mark options={draw=green, fill=green, solid}] 
	table[x expr=\thisrowno{0}, y expr=\thisrowno{1}, col sep =comma] 
	{./data/data_cubic_system4_input3.txt};
	
	
	\addplot 
	[mark=x, mark size=1.5pt, only marks, 
		mark options={draw=cyan, fill=cyan, solid}] 
	table[x expr=\thisrowno{0}, y expr=\thisrowno{1}, col sep =comma] 
	{./data/data_cubic_system4_input4.txt};
	
	
	\addplot 
	[mark=o, mark size=1.5pt, only marks, 
		mark options={draw=magenta, fill=magenta, solid}] 
	table[x expr=\thisrowno{0}, y expr=\thisrowno{1}, col sep =comma] 
	{./data/data_cubic_system4_input5.txt};
	
	
	\addplot 
	[dashed, draw=blue, line width=2pt] 
	table[x expr=\thisrowno{0}, y expr=\thisrowno{1}, col sep =comma] 
	{./data/data_cubic_system4_linear.txt};

\end{axis}
	

%% file: plot_cubic_G_system3_std.tex
	
\begin{axis}[
grid=major, 
xmin=0, xmax=0.5,	
ymin=-35, ymax=5,
width=8.5cm, height=4.00 cm, 
xlabel={\footnotesize Normalized frequency},
ylabel={\footnotesize standard dev. [dB]},
clip marker paths=true
]		
	\node at (0.475,2) {(g)};

	
	\addplot 
	[dotted, mark=*, mark size=0.5pt, 
		mark options={draw=blue,fill=blue, solid}] 
	table[x expr=\thisrowno{0}, y expr=\thisrowno{2}, col sep =comma] 
	{./data/data_cubic_system3_input1.txt};
	
	
	\addplot 
	[dotted, mark=+, mark size=1pt, 
		mark options={draw=black, fill=black, solid}] 
	table[x expr=\thisrowno{0}, y expr=\thisrowno{2}, col sep =comma] 
	{./data/data_cubic_system3_input2.txt};
	
	
	\addplot 
	[dotted, mark=square, mark size=1pt, 
		mark options={draw=green, fill=green, solid}] 
	table[x expr=\thisrowno{0}, y expr=\thisrowno{2}, col sep =comma] 
	{./data/data_cubic_system3_input3.txt};
	
	
	\addplot 
	[dotted, mark=x, mark size=1pt,
		mark options={draw=cyan, fill=cyan, solid}] 
	table[x expr=\thisrowno{0}, y expr=\thisrowno{2}, col sep =comma] 
	{./data/data_cubic_system3_input4.txt};
	
	
	\addplot 
	[dotted, mark=o, mark size=1pt,
		mark options={draw=magenta, fill=magenta, solid}] 
	table[x expr=\thisrowno{0}, y expr=\thisrowno{2}, col sep =comma] 
	{./data/data_cubic_system3_input5.txt};
	

\end{axis}
	

%% file: plot_cubic_G_system4_std.tex
	
\begin{axis}[
grid=major, 
xmin=0, xmax=0.5,	
ymin=-35, ymax=5,
width=8.5cm, height=4.00 cm, 
xlabel={\footnotesize Normalized frequency},
ylabel={\footnotesize standard dev. [dB]},
clip marker paths=true
]		
	\node at (0.475,2) {(h)};

	
	\addplot 
	[dotted, mark=*, mark size=0.5pt, 
		mark options={draw=blue,fill=blue, solid}] 
	table[x expr=\thisrowno{0}, y expr=\thisrowno{2}, col sep =comma] 
	{./data/data_cubic_system4_input1.txt};
	
	
	\addplot 
	[dotted, mark=+, mark size=1pt, 
		mark options={draw=black, fill=black, solid}] 
	table[x expr=\thisrowno{0}, y expr=\thisrowno{2}, col sep =comma] 
	{./data/data_cubic_system4_input2.txt};
	
	
	\addplot 
	[dotted, mark=square, mark size=1pt, 
		mark options={draw=green, fill=green, solid}] 
	table[x expr=\thisrowno{0}, y expr=\thisrowno{2}, col sep =comma] 
	{./data/data_cubic_system4_input3.txt};
	
	
	\addplot 
	[dotted, mark=x, mark size=1pt,
		mark options={draw=cyan, fill=cyan, solid}] 
	table[x expr=\thisrowno{0}, y expr=\thisrowno{2}, col sep =comma] 
	{./data/data_cubic_system4_input4.txt};
	
	
	\addplot 
	[dotted, mark=o, mark size=1pt,
		mark options={draw=magenta, fill=magenta, solid}] 
	table[x expr=\thisrowno{0}, y expr=\thisrowno{2}, col sep =comma] 
	{./data/data_cubic_system4_input5.txt};
	

\end{axis}
	

%% file: plot_cubic_G_ratio_system1.tex
	
\begin{axis}[
grid=major, 
xmin=0, xmax=0.5,	
ymin=-6, ymax=16,
width=8.5cm, height=4.00 cm, 
legend columns=-1,
legend to name=legend_location2,
xlabel={\footnotesize Normalized frequency},
ylabel={\footnotesize $\left|G/\hat{G}_{BLA}\right|$ [dB]},
clip marker paths=true
]	
	
	\node at (0.475,14) {(a)};

	\addplot 
	[mark=*, mark size=0.5pt, only marks, 
		mark options={draw=blue,fill=blue, solid}] 
	table[x expr=\thisrowno{0}, y expr=\thisrowno{4}, col sep =comma] 
	{./data/data_cubic_system1_input1.txt};
	\addlegendentry{DS; \ \ }
	
	\addplot 
	[mark=+, mark size=1.5pt, only marks, 
		mark options={draw=black, fill=black, solid}] 
	table[x expr=\thisrowno{0}, y expr=\thisrowno{4}, col sep =comma] 
	{./data/data_cubic_system1_input2.txt};
	\addlegendentry{RCS;  \ \ }	
	
	\addplot 
	[mark=square, mark size=1.5pt, only marks, 
		mark options={draw=green, fill=green, solid}] 
	table[x expr=\thisrowno{0}, y expr=\thisrowno{4}, col sep =comma] 
	{./data/data_cubic_system1_input3.txt};
	\addlegendentry{WGN;  \ \ }	
	
	\addplot 
	[mark=x, mark size=1.5pt, only marks, 
		mark options={draw=cyan, fill=cyan, solid}] 
	table[x expr=\thisrowno{0}, y expr=\thisrowno{4}, col sep =comma] 
	{./data/data_cubic_system1_input4.txt};
	\addlegendentry{MLBS;  \ \ }		
	
	\addplot 
	[mark=o, mark size=1.5pt, only marks, 
		mark options={draw=magenta, fill=magenta, solid}] 
	table[x expr=\thisrowno{0}, y expr=\thisrowno{4}, col sep =comma] 
	{./data/data_cubic_system1_input5.txt};
	\addlegendentry{IRMLBS; \ \ }
	
	\addplot 
	[dashed, draw=blue, line width=2pt, samples=2, domain=0:0.5] {0};
	\addlegendentry{Linear;}

\end{axis}
	

%% file: plot_cubic_G_ratio_system2.tex
	
\begin{axis}[
grid=major, 
xmin=0, xmax=0.5,	
ymin=-6, ymax=16,
width=8.5cm, height=4.00 cm, 
xlabel={\footnotesize Normalized frequency},
ylabel={\footnotesize $\left|G/\hat{G}_{BLA}\right|$ [dB]},
clip marker paths=true
]	
	
	\node at (0.475,14) {(b)};

	\addplot 
	[mark=*, mark size=0.5pt, only marks, 
		mark options={draw=blue,fill=blue, solid}] 
	table[x expr=\thisrowno{0}, y expr=\thisrowno{4}, col sep =comma] 
	{./data/data_cubic_system2_input1.txt};
	
	\addplot 
	[mark=+, mark size=1.5pt, only marks, 
		mark options={draw=black, fill=black, solid}] 
	table[x expr=\thisrowno{0}, y expr=\thisrowno{4}, col sep =comma] 
	{./data/data_cubic_system2_input2.txt};
	
	\addplot 
	[mark=square, mark size=1.5pt, only marks, 
		mark options={draw=green, fill=green, solid}] 
	table[x expr=\thisrowno{0}, y expr=\thisrowno{4}, col sep =comma] 
	{./data/data_cubic_system2_input3.txt};
	
	\addplot 
	[mark=x, mark size=1.5pt, only marks, 
		mark options={draw=cyan, fill=cyan, solid}] 
	table[x expr=\thisrowno{0}, y expr=\thisrowno{4}, col sep =comma] 
	{./data/data_cubic_system2_input4.txt};
	
	\addplot 
	[mark=o, mark size=1.5pt, only marks, 
		mark options={draw=magenta, fill=magenta, solid}] 
	table[x expr=\thisrowno{0}, y expr=\thisrowno{4}, col sep =comma] 
	{./data/data_cubic_system2_input5.txt};
	
	\addplot 
	[dashed, draw=blue, line width=2pt, samples=2, domain=0:0.5] {0};

\end{axis}
	

%% file: plot_cubic_G_ratio_system3.tex
	
\begin{axis}[
grid=major, 
xmin=0, xmax=0.5,	
ymin=-6, ymax=16,
width=8.5cm, height=4.00 cm, 
xlabel={\footnotesize Normalized frequency},
ylabel={\footnotesize $\left|G/\hat{G}_{BLA}\right|$ [dB]},
clip marker paths=true
]	
	
	\node at (0.475,14) {(c)};

	\addplot 
	[mark=*, mark size=0.5pt, only marks, 
		mark options={draw=blue,fill=blue, solid}] 
	table[x expr=\thisrowno{0}, y expr=\thisrowno{4}, col sep =comma] 
	{./data/data_cubic_system3_input1.txt};
	
	\addplot 
	[mark=+, mark size=1.5pt, only marks, 
		mark options={draw=black, fill=black, solid}] 
	table[x expr=\thisrowno{0}, y expr=\thisrowno{4}, col sep =comma] 
	{./data/data_cubic_system3_input2.txt};
	
	\addplot 
	[mark=square, mark size=1.5pt, only marks, 
		mark options={draw=green, fill=green, solid}] 
	table[x expr=\thisrowno{0}, y expr=\thisrowno{4}, col sep =comma] 
	{./data/data_cubic_system3_input3.txt};
	
	\addplot 
	[mark=x, mark size=1.5pt, only marks, 
		mark options={draw=cyan, fill=cyan, solid}] 
	table[x expr=\thisrowno{0}, y expr=\thisrowno{4}, col sep =comma] 
	{./data/data_cubic_system3_input4.txt};
	
	\addplot 
	[mark=o, mark size=1.5pt, only marks, 
		mark options={draw=magenta, fill=magenta, solid}] 
	table[x expr=\thisrowno{0}, y expr=\thisrowno{4}, col sep =comma] 
	{./data/data_cubic_system3_input5.txt};
	
	\addplot 
	[dashed, draw=blue, line width=2pt, samples=2, domain=0:0.5] {0};

\end{axis}
	

%% file: plot_cubic_G_ratio_system4.tex
	
\begin{axis}[
grid=major, 
xmin=0, xmax=0.5,	
ymin=-6, ymax=16,
width=8.5cm, height=4.00 cm, 
xlabel={\footnotesize Normalized frequency},
ylabel={\footnotesize $\left|G/\hat{G}_{BLA}\right|$ [dB]},
clip marker paths=true
]	
	
	\node at (0.475,14) {(d)};

	\addplot 
	[mark=*, mark size=0.5pt, only marks, 
		mark options={draw=blue,fill=blue, solid}] 
	table[x expr=\thisrowno{0}, y expr=\thisrowno{4}, col sep =comma] 
	{./data/data_cubic_system4_input1.txt};
	
	\addplot 
	[mark=+, mark size=1.5pt, only marks, 
		mark options={draw=black, fill=black, solid}] 
	table[x expr=\thisrowno{0}, y expr=\thisrowno{4}, col sep =comma] 
	{./data/data_cubic_system4_input2.txt};
	
	\addplot 
	[mark=square, mark size=1.5pt, only marks, 
		mark options={draw=green, fill=green, solid}] 
	table[x expr=\thisrowno{0}, y expr=\thisrowno{4}, col sep =comma] 
	{./data/data_cubic_system4_input3.txt};
	
	\addplot 
	[mark=x, mark size=1.5pt, only marks, 
		mark options={draw=cyan, fill=cyan, solid}] 
	table[x expr=\thisrowno{0}, y expr=\thisrowno{4}, col sep =comma] 
	{./data/data_cubic_system4_input4.txt};
	
	\addplot 
	[mark=o, mark size=1.5pt, only marks, 
		mark options={draw=magenta, fill=magenta, solid}] 
	table[x expr=\thisrowno{0}, y expr=\thisrowno{4}, col sep =comma] 
	{./data/data_cubic_system4_input5.txt};
	
	\addplot 
	[dashed, draw=blue, line width=2pt, samples=2, domain=0:0.5] {0};

\end{axis}
	

%% file: Fig_4.tex
%
%
\definecolor{mycolor1}{rgb}{0.00000,0.44700,0.74100}%
\begin{tikzpicture}

\begin{axis}[%
width=8.5cm, height=4.00 cm, 
xmin=-10,
xmax=762,
xlabel style={font=\color{white!15!black}},
xlabel={$r$},
ymin=-1.7,
ymax=1.7,
ylabel style={font=\color{white!15!black}},
ylabel={$R_{yu}(r)$},
axis background/.style={fill=white},
legend style={legend cell align=left, align=left, draw=white!15!black},
legend style={at={(0.5,1.05)}, anchor=south},
legend columns=-1,
clip marker paths=true
]
\addplot[ycomb, color=mycolor1, mark=o, mark options={solid, mycolor1}] table[row sep=crcr] {%
1	1.43767884514436\\
2	1.24129936399474\\
3	0.611368053947369\\
4	-0.00166503557312254\\
5	-0.00410822163588391\\
6	-0.00283719550858656\\
7	-0.00158350925925927\\
8	3.93668874172194e-05\\
9	-0.00067454907161803\\
10	0.00320862682602923\\
11	0.00200484308510637\\
12	-0.00193467376830892\\
13	0.00433154266666666\\
14	-0.0011467917222964\\
15	-0.000385978609625673\\
16	0.00155006827309237\\
17	0.0021700308310992\\
18	0.0016697932885906\\
19	0.000421446236559149\\
20	0.000669209959623148\\
21	0.000340816711590292\\
22	0.0027527935222672\\
23	0.00197532432432432\\
24	0.00116117456021651\\
25	0.000106703252032521\\
26	-0.00237221031207598\\
27	-0.00571430706521739\\
28	0.00159994557823129\\
29	5.11525885558563e-05\\
30	-0.00103085129604367\\
31	-0.00381671584699453\\
32	-0.00148707660738715\\
33	-7.22328767123303e-05\\
34	-0.00236582990397806\\
35	-0.00273410439560439\\
36	-0.00228673590096286\\
37	-0.00185402341597796\\
38	-0.000853689655172411\\
39	0.00133490055248619\\
40	-0.000313658367911476\\
41	0.00122627700831025\\
42	0.000759233009708733\\
43	0.000551852777777782\\
44	0.000860573018080668\\
45	0.000995725626740951\\
46	-0.000952670850767084\\
47	0.0013517094972067\\
48	-0.00335721398601399\\
49	-0.000922085434173668\\
50	0.00111759467040674\\
51	0.000415087078651681\\
52	-0.0016988523206751\\
53	-0.0027642014084507\\
54	-0.00247824400564174\\
55	-0.00315546610169492\\
56	0.00346489250353607\\
57	0.00281811898016997\\
58	0.00225582411347517\\
59	-0.000848291193181816\\
60	0.00142598577524894\\
61	-0.000737400284900287\\
62	-0.000662587731811686\\
63	0.00317831428571429\\
64	0.00366534620886981\\
65	-0.00118530229226361\\
66	-0.00171679196556673\\
67	0.00439571120689654\\
68	-0.00107412230215827\\
69	0.00160223198847262\\
70	0.00103703751803752\\
71	0.00200896676300579\\
72	0.000392218523878435\\
73	-0.000274104347826088\\
74	0.00129853120464441\\
75	-0.0003690523255814\\
76	-0.00280740029112082\\
77	0.00254222594752187\\
78	-0.000281623357664236\\
79	-0.000491476608187134\\
80	0.00144084187408492\\
81	0.00262730938416422\\
82	0.000779209985315712\\
83	-0.000769692647058822\\
84	-0.000722050073637702\\
85	-0.000995920353982303\\
86	-0.00107798375184638\\
87	0.000872140532544375\\
88	-0.000686096296296298\\
89	-0.000706075667655788\\
90	0.00362923625557207\\
91	0.00349841517857143\\
92	0.000347488822652761\\
93	-0.00419408805970149\\
94	-0.000836992526158446\\
95	-0.00443629041916167\\
96	-0.000221244377811098\\
97	-0.00379217417417417\\
98	-0.00206466165413534\\
99	0.000684963855421689\\
100	0.0030056923076923\\
101	-0.000656419939577046\\
102	-0.000331632375189102\\
103	0.000222395454545457\\
104	-0.0013312564491654\\
105	0.00243348176291793\\
106	-0.000609729071537293\\
107	-0.000501306402439024\\
108	0.000203170992366413\\
109	-0.00253200152905199\\
110	-0.000592197549770289\\
111	-0.000908401840490799\\
112	0.00234427956989247\\
113	0.00342084307692308\\
114	-0.00261426040061633\\
115	-0.00393361419753087\\
116	-0.00170583925811438\\
117	-0.00486496439628484\\
118	-0.00216649612403101\\
119	-0.000727159937888195\\
120	1.02208398133789e-05\\
121	-0.00152517289719626\\
122	0.000899031201248052\\
123	-0.0022073265625\\
124	-0.000776910798122084\\
125	-0.00162886677115987\\
126	-0.000751470957613836\\
127	-0.00378380974842767\\
128	0.71761122047244\\
129	0.621996973186119\\
130	0.309183017377567\\
131	0.00388702215189874\\
132	0.0010518969889065\\
133	0.00257143333333335\\
134	-0.00292146740858506\\
135	0.000639320063694269\\
136	0.00120188038277513\\
137	0.00353415495207668\\
138	0.0010939776\\
139	3.67676282051262e-05\\
140	0.0017810658105939\\
141	-0.00113203536977493\\
142	0.006344576489533\\
143	0.00518927580645161\\
144	0.00140327948303716\\
145	-0.0019498786407767\\
146	0.00222919124797406\\
147	0.00173982142857142\\
148	0.00207026666666667\\
149	-0.00164997068403909\\
150	0.0026952854812398\\
151	0.00107824346405229\\
152	0.00437924713584289\\
153	-0.000502321311475411\\
154	-0.00167490147783251\\
155	0.00225547861842105\\
156	-0.000544630971993411\\
157	-0.00192980693069307\\
158	-0.00243559504132232\\
159	-0.00327348178807947\\
160	-0.00365722553897181\\
161	-0.000500709302325582\\
162	-1.44608985025095e-05\\
163	-0.000748536666666669\\
164	-0.00197634223706177\\
165	0.00195364548494983\\
166	0.000860433835845899\\
167	-0.00060551677852349\\
168	0.00213954453781512\\
169	0.00168398148148148\\
170	0.0021277892074199\\
171	0.00119531418918919\\
172	-0.00198215566835871\\
173	-0.00360231016949153\\
174	-0.000706402376910017\\
175	-0.0037625306122449\\
176	-0.0018984940374787\\
177	0.000250663822525597\\
178	0.00174664786324787\\
179	-0.000321008561643834\\
180	-0.00160841852487135\\
181	0.000285649484536085\\
182	-0.00134584681583477\\
183	0.00340420689655173\\
184	0.00305270639032815\\
185	0.00248304671280277\\
186	0.000836611785095319\\
187	0.000728111111111111\\
188	0.00137635130434783\\
189	0.00285117073170732\\
190	0.00204261431064574\\
191	0.00111176223776224\\
192	-0.0011981401050788\\
193	0.000497833333333327\\
194	0.00164954130052723\\
195	-0.00013986971830987\\
196	0.00128409876543209\\
197	0.00233464664310954\\
198	0.00281650973451327\\
199	-0.000113918439716311\\
200	-0.000117657193605681\\
201	-0.0014698024911032\\
202	-0.00211997147950089\\
203	-0.00179505892857143\\
204	0.00434243291592129\\
205	-0.00101708781362007\\
206	-0.000179859964093356\\
207	0.000899280575539574\\
208	0.00288078378378379\\
209	0.00199675451263538\\
210	-0.00366043580470162\\
211	0.000709253623188404\\
212	0.00162036297640653\\
213	0.00214070363636363\\
214	0.00257607103825136\\
215	0.00266344160583942\\
216	-0.00155751188299817\\
217	-0.000820738095238096\\
218	0.0029367504587156\\
219	0.00502769485294118\\
220	-0.00174984346224678\\
221	-0.00129186715867159\\
222	-0.00378506839186691\\
223	-0.000321168518518512\\
224	-0.000192068645640073\\
225	-0.00269899814126394\\
226	0.00267224022346369\\
227	0.00261530410447761\\
228	-0.00148304112149533\\
229	0.00127857303370787\\
230	0.00231395684803002\\
231	0.00260093609022556\\
232	-0.00067082674199624\\
233	0.0013897679245283\\
234	0.000906378071833641\\
235	0.00286769507575758\\
236	5.43130929791331e-05\\
237	0.00136919011406844\\
238	0.00067917904761905\\
239	0.00363658778625953\\
240	0.00145891395793499\\
241	-0.00150000383141762\\
242	-0.0052508368522073\\
243	-0.00030957884615384\\
244	-0.00384385549132949\\
245	-6.62548262547874e-06\\
246	0.000109949709864603\\
247	0.00128924418604653\\
248	0.000919784466019412\\
249	0.00391169455252918\\
250	-0.00348842884990254\\
251	-0.001408875\\
252	-0.0001197436399217\\
253	0.00265991176470588\\
254	0.0031873379174853\\
255	-0.721359572834646\\
256	-0.620713098619329\\
257	-0.302959122529644\\
258	0.00508097425742577\\
259	0.00626447619047621\\
260	0.00623684691848909\\
261	-0.00118535258964142\\
262	0.00121179640718563\\
263	0.001180324\\
264	-0.000412649298597206\\
265	-0.00169943775100402\\
266	0.00139819718309859\\
267	-0.00266160887096774\\
268	-0.000226565656565649\\
269	0.00645754453441296\\
270	0.00232407910750507\\
271	-0.000808648373983751\\
272	-0.00312521995926681\\
273	0.00269716326530612\\
274	0.00210448261758692\\
275	0.00300698770491804\\
276	-0.00440166940451744\\
277	0.000420810699588476\\
278	-0.000455232989690721\\
279	0.00415785123966942\\
280	0.000709968944099378\\
281	0.00395890871369295\\
282	0.000898987525987521\\
283	-0.00145956666666667\\
284	0.000359292275574113\\
285	0.00140233891213389\\
286	-0.00202708176100628\\
287	-0.00380567016806722\\
288	0.00182236\\
289	0.00235236075949367\\
290	0.000948281183932336\\
291	0.0014825720338983\\
292	0.00256346709129511\\
293	0.00174964468085106\\
294	5.59424307036265e-05\\
295	0.000146130341880342\\
296	0.001251426124197\\
297	0.00105057296137338\\
298	-0.000164094623655919\\
299	-0.00281195043103448\\
300	-0.0024811879049676\\
301	-0.00154681601731602\\
302	-0.000957492407809107\\
303	-0.00156874130434783\\
304	-0.00127331808278867\\
305	0.000898026200873362\\
306	0.00200674179431072\\
307	0.000561557017543859\\
308	0.00331392307692308\\
309	0.00213625550660794\\
310	0.000989962472406179\\
311	0.00139936283185841\\
312	-8.22239467849216e-05\\
313	0.00161732666666667\\
314	-0.00163737639198219\\
315	0.00185655803571429\\
316	0.00291506935123042\\
317	-0.000529672645739906\\
318	-0.00225189213483146\\
319	0.000139590090090072\\
320	0.00134553273137696\\
321	-0.00353215610859729\\
322	0.000433469387755093\\
323	0.000523906818181816\\
324	0.00122366059225512\\
325	-0.000288164383561644\\
326	-0.0011396590389016\\
327	-0.000659339449541285\\
328	-0.00406757931034483\\
329	-0.00155644930875576\\
330	0.00117424711316397\\
331	0.00184293981481481\\
332	-0.000494373549883988\\
333	0.00109542558139535\\
334	-0.00154903263403263\\
335	0.00104297897196262\\
336	0.0020986206088993\\
337	-0.00120176056338028\\
338	0.000434136470588233\\
339	0.00343806132075472\\
340	0.00411733333333334\\
341	0.00180464454976304\\
342	0.00400711401425179\\
343	-0.000944023809523809\\
344	-0.00517554415274462\\
345	-0.00224031578947369\\
346	0.00265040527577937\\
347	0.00219364903846154\\
348	-0.000510568674698794\\
349	0.00137755072463768\\
350	-0.000777334140435824\\
351	0.00298591019417475\\
352	-0.00110785158150853\\
353	0.000971939024390243\\
354	0.000343528117359431\\
355	0.000258485294117645\\
356	0.00249876904176905\\
357	0.00127479064039409\\
358	0.00286426666666668\\
359	-0.00167353712871287\\
360	0.00119849379652605\\
361	0.00128155223880597\\
362	0.00255095511221945\\
363	0.00202073\\
364	0.00283056140350878\\
365	0.00201639447236181\\
366	0.000160465994962215\\
367	-0.00202881565656568\\
368	0.00230612911392405\\
369	-0.00114793401015229\\
370	0.0016459541984733\\
371	0.00158547704081633\\
372	0.00255749104859335\\
373	0.00162969743589743\\
374	0.000863051413881707\\
375	0.00213893298969075\\
376	0.00294599224806201\\
377	-0.00129957512953368\\
378	-0.0016189688311688\\
379	0.00200799739583334\\
380	0.00266046475195825\\
381	0.00613865706806286\\
382	-1.43767884514436\\
383	-1.24134315526316\\
384	-0.611452\\
385	0.00157671957671958\\
386	0.00405581697612735\\
387	0.00312369148936172\\
388	0.0014935466666667\\
389	-0.000276053475935814\\
390	-0.000363466487935643\\
391	-0.00350180645161289\\
392	-0.00230572237196767\\
393	0.00175870270270271\\
394	-0.00437149593495935\\
395	0.0013403695652174\\
396	0.000802231607629439\\
397	-0.00141628961748634\\
398	-0.00198261643835617\\
399	-0.0010452197802198\\
400	-0.000568134986225895\\
401	-0.00103935082872928\\
402	-0.000948867036011066\\
403	-0.00335283333333333\\
404	-0.00179474373259053\\
405	-0.00072985754189944\\
406	-0.000241383753501414\\
407	0.00253171348314608\\
408	0.0056903690140845\\
409	-0.00178103107344633\\
410	-0.000496090651558072\\
411	0.000853238636363634\\
412	0.00419080911680912\\
413	0.00214521142857143\\
414	0.000618699140401146\\
415	0.0031181235632184\\
416	0.00302893083573488\\
417	0.00254183815028902\\
418	0.00195699130434781\\
419	0.00112399418604651\\
420	-0.000896868804664732\\
421	0.00132250584795321\\
422	-0.00112305865102639\\
423	-0.00113222352941177\\
424	-0.00137979056047198\\
425	-0.00132457100591716\\
426	-0.00109695548961425\\
427	0.00113220535714285\\
428	-0.00142750746268656\\
429	0.00356353892215569\\
430	0.00124548048048048\\
431	-0.00145070783132529\\
432	-0.000270583081571\\
433	0.00271434545454546\\
434	0.00314579635258359\\
435	0.00343776524390244\\
436	0.00370899082568808\\
437	-0.00444185889570552\\
438	-0.00332972\\
439	-0.00206784259259259\\
440	0.000759510835913315\\
441	-0.00192038819875777\\
442	0.000565227414330214\\
443	-7.13062499999996e-05\\
444	-0.00408160188087774\\
445	-0.00301656289308176\\
446	0.00216334384858045\\
447	0.0013619240506329\\
448	-0.00492889841269841\\
449	0.000595872611464972\\
450	-0.00160908626198082\\
451	-0.00152550961538463\\
452	-0.00320150160771704\\
453	-0.00150126451612904\\
454	3.70809061488658e-05\\
455	-0.00167321103896104\\
456	0.000152286644951132\\
457	0.00255963725490196\\
458	-0.00429481967213114\\
459	-0.000807098684210526\\
460	0.000554821782178223\\
461	-0.00132269867549669\\
462	-0.00294547176079734\\
463	-0.000574503333333338\\
464	0.00245913377926421\\
465	0.00216015436241611\\
466	0.00157204377104377\\
467	0.00099709797297298\\
468	-0.000119722033898303\\
469	0.00156193197278911\\
470	-0.000670443686006825\\
471	-0.0058800993150685\\
472	-0.0052339793814433\\
473	-0.00173477586206897\\
474	0.00514632179930796\\
475	0.000169406249999998\\
476	0.00406325087108014\\
477	-0.000283108391608391\\
478	0.00473662105263156\\
479	0.00395318661971831\\
480	-0.00155384452296819\\
481	-0.00480107446808512\\
482	0.000473754448398583\\
483	-0.000161553571428573\\
484	-0.000617906810035828\\
485	0.000540902877697846\\
486	-0.00291317328519853\\
487	-7.627173913044e-05\\
488	0.000161305454545455\\
489	0.00011022262773723\\
490	0.00218315750915751\\
491	-0.00181750735294118\\
492	0.000334225092250929\\
493	-0.00371001111111109\\
494	-0.00440565799256505\\
495	0.00254830970149254\\
496	0.00417666666666665\\
497	0.00267855639097744\\
498	0.00467228301886793\\
499	0.00192856060606061\\
500	0.00110878707224334\\
501	0.00240754580152672\\
502	0.00375621455938696\\
503	0.00152046153846155\\
504	0.00463568725868725\\
505	0.00209909689922481\\
506	-5.81984435797577e-05\\
507	5.16874999999963e-05\\
508	0.00385022352941175\\
509	-0.71747981496063\\
510	-0.623521245059288\\
511	-0.310047265873016\\
512	-0.00536001593625503\\
513	-0.000553744000000004\\
514	-0.000605586345381535\\
515	0.00578875806451612\\
516	0.000228356275303666\\
517	-0.00216526016260162\\
518	-0.00356195102040815\\
519	-0.00144328688524589\\
520	-4.90864197530862e-05\\
521	-0.00148026033057851\\
522	0.00152656016597511\\
523	-0.00621002083333332\\
524	-0.00491635983263598\\
525	-0.0017566050420168\\
526	0.00262997046413502\\
527	-0.00348283050847459\\
528	-0.00372712765957446\\
529	-0.00309862393162391\\
530	0.00104599570815452\\
531	-0.00392065517241379\\
532	1.34372294372319e-05\\
533	-0.0052605\\
534	-0.00123360698689957\\
535	9.4903508771937e-05\\
536	-0.00213439207048458\\
537	0.000681106194690265\\
538	0.000631586666666667\\
539	0.00234866964285714\\
540	0.00550321524663677\\
541	0.00715988738738739\\
542	0.00273875565610861\\
543	-0.00277399545454545\\
544	0.000897890410958906\\
545	0.00492297706422017\\
546	-0.0040773456221198\\
547	-0.00115022685185183\\
548	-0.00071712558139534\\
549	-0.00385606074766356\\
550	-0.00336686854460094\\
551	-0.00360068396226415\\
552	-0.00363890047393365\\
553	0.00229519999999999\\
554	0.00578055023923444\\
555	0.000292793269230769\\
556	0.00444103381642514\\
557	0.00102049514563106\\
558	-0.00258772682926829\\
559	-0.00375749999999999\\
560	0.000238221674876847\\
561	0.000377995049504943\\
562	-0.000615502487562191\\
563	0.00131799999999999\\
564	-0.00536004020100503\\
565	-0.00311918181818182\\
566	-0.00477909137055837\\
567	-0.00271185714285715\\
568	-0.00396025641025641\\
569	-0.00252708762886598\\
570	-0.00598690155440414\\
571	-0.00457026041666667\\
572	-0.00190898952879581\\
573	0.00379836315789475\\
574	0.000186883597883607\\
575	-0.00261846808510635\\
576	-0.00302839037433154\\
577	-0.00073732258064517\\
578	-0.00336636756756759\\
579	-0.00619865760869562\\
580	-5.02076502732064e-05\\
581	0.000117626373626366\\
582	-0.000692193370165738\\
583	0.00304289444444445\\
584	0.00307998324022346\\
585	-0.00392892696629214\\
586	0.00191966666666667\\
587	0.000169085227272726\\
588	-0.00444618285714285\\
589	-0.00253496551724137\\
590	0.000388398843930634\\
591	0.00903548255813952\\
592	0.00141115789473684\\
593	-0.00135596470588235\\
594	-0.00295447928994082\\
595	-0.00513256547619048\\
596	-0.00255026347305389\\
597	-0.00113351204819278\\
598	-0.000754284848484859\\
599	-0.00639266463414634\\
600	-0.00988876073619632\\
601	0.00575961111111111\\
602	0.0011984099378882\\
603	0.0022619125\\
604	-0.000342283018867922\\
605	0.00401025949367088\\
606	0.00796545859872615\\
607	-0.00566536538461538\\
608	-0.00522440645161291\\
609	0.00126854545454544\\
610	-0.00162294117647062\\
611	-0.00262868421052632\\
612	-0.00427499337748343\\
613	0.00260225333333333\\
614	-0.00344603355704697\\
615	-0.00230706081081081\\
616	-0.00471251700680272\\
617	-0.0048314589041096\\
618	-0.00602167586206898\\
619	-0.0027176875\\
620	-0.0083883006993007\\
621	-0.00411009154929579\\
622	0.00452236879432625\\
623	0.00828224285714287\\
624	0.00469556834532375\\
625	0.00509123188405796\\
626	0.000322080291970799\\
627	0.00212872794117646\\
628	-0.000552614814814826\\
629	-0.00146961940298507\\
630	-0.00284693233082705\\
631	0.00776105303030302\\
632	0.00319240458015265\\
633	-0.00145383076923078\\
634	-0.00664619379844965\\
635	-0.00585443750000001\\
636	0.725498228346458\\
637	0.622050238095238\\
638	0.303830736\\
639	-0.00634593548387109\\
640	-0.00851216260162615\\
641	-0.00569404098360677\\
642	0.0068604049586775\\
643	-0.000630758333333399\\
644	0.00231047058823526\\
645	0.00366263559322026\\
646	0.00449294871794872\\
647	0.00102009482758621\\
648	0.00379099130434781\\
649	0.00124224561403508\\
650	-0.00660315929203539\\
651	0.00229491964285713\\
652	-0.000421396396396384\\
653	0.000811136363636358\\
654	-0.00803183486238531\\
655	-0.00883532407407408\\
656	-0.00775344859813086\\
657	0.00510799999999999\\
658	-0.00266021904761903\\
659	0.00275921153846154\\
660	-0.00523464077669903\\
661	-0.000831352941176468\\
662	-0.00713784158415842\\
663	-0.000910639999999991\\
664	0.00668186868686868\\
665	-0.00759370408163265\\
666	-0.00304009278350516\\
667	0.00576230208333334\\
668	0.0108683263157895\\
669	0.000827659574468088\\
670	-0.00854408602150535\\
671	0.000890869565217386\\
672	-0.00153335164835168\\
673	-0.0076847333333333\\
674	-0.0136623033707865\\
675	-0.00871396590909088\\
676	-0.00141340229885059\\
677	-0.00539215116279071\\
678	0.000932729411764784\\
679	-0.00184434523809522\\
680	0.00325332530120482\\
681	0.00651470731707316\\
682	-0.0016067901234568\\
683	0.00448145000000001\\
684	0.000772797468354447\\
685	-0.00146605128205128\\
686	-0.00470150649350647\\
687	-0.00979001315789472\\
688	-0.00253076000000002\\
689	-0.0112947972972973\\
690	-0.00639680821917808\\
691	-0.00753186111111114\\
692	-0.00549332394366198\\
693	-0.00556531428571429\\
694	-0.0061553768115942\\
695	-2.61617647058864e-05\\
696	-0.00290110447761195\\
697	-0.00507036363636364\\
698	-0.00605579999999998\\
699	-0.00514698437499999\\
700	0.00185926984126985\\
701	0.00789114516129035\\
702	0.00820849180327875\\
703	-0.0046250666666666\\
704	-0.00420138983050847\\
705	-0.00140108620689654\\
706	0.00320092982456146\\
707	0.0110528928571429\\
708	0.00764205454545464\\
709	0.00812485185185185\\
710	0.00454852830188682\\
711	0.00352057692307694\\
712	0.00976223529411762\\
713	0.00847181999999999\\
714	-0.00762034693877551\\
715	-0.00412889583333335\\
716	-0.00412514893617016\\
717	-0.00176102173913044\\
718	-0.00377906666666666\\
719	0.00670122727272727\\
720	-0.0136533720930233\\
721	-0.0149737619047619\\
722	-0.0186296829268293\\
723	-0.0162354\\
724	0.00073597435897435\\
725	0.0229075263157894\\
726	0.0178577567567567\\
727	0.00799719444444445\\
728	0.0112310285714286\\
729	0.00634697058823528\\
730	-0.0139040303030303\\
731	0.010053875\\
732	0.0153403870967743\\
733	0.0168046000000002\\
734	0.00474562068965523\\
735	-0.00669121428571433\\
736	-0.0148928888888888\\
737	-0.0122348846153846\\
738	0.0136731600000001\\
739	0.0127113749999999\\
740	-0.00365408695652166\\
741	0.00735309090909088\\
742	-0.00441223809523798\\
743	-0.0182404\\
744	-0.0223755263157895\\
745	-0.0222931111111112\\
746	-0.0197411764705883\\
747	0.00736981249999999\\
748	-0.00370906666666659\\
749	-0.00304007142857134\\
750	0.0238404615384617\\
751	0.0196687500000002\\
752	-0.000842363636363598\\
753	-0.00710359999999994\\
754	-0.0125056666666667\\
755	-0.0478036250000002\\
756	-0.0791938571428573\\
757	-0.0799548333333333\\
758	-0.0102845999999997\\
759	0.0687607499999998\\
760	0.0109720000000001\\
761	-0.0271535000000004\\
762	-0.0472840000000012\\
};
\addplot[forget plot, color=white!15!black] table[row sep=crcr] {%
0	0\\
762	0\\
};
\addlegendentry{simulation}

\addplot[ycomb, color=black, mark=x, mark options={solid, black}] table[row sep=crcr] {%
1	1.44\\
2	1.246\\
3	0.614\\
4	0\\
5	0\\
6	0\\
7	0\\
8	0\\
9	0\\
10	0\\
11	0\\
12	0\\
13	0\\
14	0\\
15	0\\
16	0\\
17	0\\
18	0\\
19	0\\
20	0\\
21	0\\
22	0\\
23	0\\
24	0\\
25	0\\
26	0\\
27	0\\
28	0\\
29	0\\
30	0\\
31	0\\
32	0\\
33	0\\
34	0\\
35	0\\
36	0\\
37	0\\
38	0\\
39	0\\
40	0\\
41	0\\
42	0\\
43	0\\
44	0\\
45	0\\
46	0\\
47	0\\
48	0\\
49	0\\
50	0\\
51	0\\
52	0\\
53	0\\
54	0\\
55	0\\
56	0\\
57	0\\
58	0\\
59	0\\
60	0\\
61	0\\
62	0\\
63	0\\
64	0\\
65	0\\
66	0\\
67	0\\
68	0\\
69	0\\
70	0\\
71	0\\
72	0\\
73	0\\
74	0\\
75	0\\
76	0\\
77	0\\
78	0\\
79	0\\
80	0\\
81	0\\
82	0\\
83	0\\
84	0\\
85	0\\
86	0\\
87	0\\
88	0\\
89	0\\
90	0\\
91	0\\
92	0\\
93	0\\
94	0\\
95	0\\
96	0\\
97	0\\
98	0\\
99	0\\
100	0\\
101	0\\
102	0\\
103	0\\
104	0\\
105	0\\
106	0\\
107	0\\
108	0\\
109	0\\
110	0\\
111	0\\
112	0\\
113	0\\
114	0\\
115	0\\
116	0\\
117	0\\
118	0\\
119	0\\
120	0\\
121	0\\
122	0\\
123	0\\
124	0\\
125	0\\
126	0\\
127	0\\
128	0.72\\
129	0.623\\
130	0.307\\
131	0\\
132	0\\
133	0\\
134	0\\
135	0\\
136	0\\
137	0\\
138	0\\
139	0\\
140	0\\
141	0\\
142	0\\
143	0\\
144	0\\
145	0\\
146	0\\
147	0\\
148	0\\
149	0\\
150	0\\
151	0\\
152	0\\
153	0\\
154	0\\
155	0\\
156	0\\
157	0\\
158	0\\
159	0\\
160	0\\
161	0\\
162	0\\
163	0\\
164	0\\
165	0\\
166	0\\
167	0\\
168	0\\
169	0\\
170	0\\
171	0\\
172	0\\
173	0\\
174	0\\
175	0\\
176	0\\
177	0\\
178	0\\
179	0\\
180	0\\
181	0\\
182	0\\
183	0\\
184	0\\
185	0\\
186	0\\
187	0\\
188	0\\
189	0\\
190	0\\
191	0\\
192	0\\
193	0\\
194	0\\
195	0\\
196	0\\
197	0\\
198	0\\
199	0\\
200	0\\
201	0\\
202	0\\
203	0\\
204	0\\
205	0\\
206	0\\
207	0\\
208	0\\
209	0\\
210	0\\
211	0\\
212	0\\
213	0\\
214	0\\
215	0\\
216	0\\
217	0\\
218	0\\
219	0\\
220	0\\
221	0\\
222	0\\
223	0\\
224	0\\
225	0\\
226	0\\
227	0\\
228	0\\
229	0\\
230	0\\
231	0\\
232	0\\
233	0\\
234	0\\
235	0\\
236	0\\
237	0\\
238	0\\
239	0\\
240	0\\
241	0\\
242	0\\
243	0\\
244	0\\
245	0\\
246	0\\
247	0\\
248	0\\
249	0\\
250	0\\
251	0\\
252	0\\
253	0\\
254	0\\
255	-0.72\\
256	-0.623\\
257	-0.307\\
258	0\\
259	0\\
260	0\\
261	0\\
262	0\\
263	0\\
264	0\\
265	0\\
266	0\\
267	0\\
268	0\\
269	0\\
270	0\\
271	0\\
272	0\\
273	0\\
274	0\\
275	0\\
276	0\\
277	0\\
278	0\\
279	0\\
280	0\\
281	0\\
282	0\\
283	0\\
284	0\\
285	0\\
286	0\\
287	0\\
288	0\\
289	0\\
290	0\\
291	0\\
292	0\\
293	0\\
294	0\\
295	0\\
296	0\\
297	0\\
298	0\\
299	0\\
300	0\\
301	0\\
302	0\\
303	0\\
304	0\\
305	0\\
306	0\\
307	0\\
308	0\\
309	0\\
310	0\\
311	0\\
312	0\\
313	0\\
314	0\\
315	0\\
316	0\\
317	0\\
318	0\\
319	0\\
320	0\\
321	0\\
322	0\\
323	0\\
324	0\\
325	0\\
326	0\\
327	0\\
328	0\\
329	0\\
330	0\\
331	0\\
332	0\\
333	0\\
334	0\\
335	0\\
336	0\\
337	0\\
338	0\\
339	0\\
340	0\\
341	0\\
342	0\\
343	0\\
344	0\\
345	0\\
346	0\\
347	0\\
348	0\\
349	0\\
350	0\\
351	0\\
352	0\\
353	0\\
354	0\\
355	0\\
356	0\\
357	0\\
358	0\\
359	0\\
360	0\\
361	0\\
362	0\\
363	0\\
364	0\\
365	0\\
366	0\\
367	0\\
368	0\\
369	0\\
370	0\\
371	0\\
372	0\\
373	0\\
374	0\\
375	0\\
376	0\\
377	0\\
378	0\\
379	0\\
380	0\\
381	0\\
382	-1.44\\
383	-1.246\\
384	-0.614\\
385	0\\
386	0\\
387	0\\
388	0\\
389	0\\
390	0\\
391	0\\
392	0\\
393	0\\
394	0\\
395	0\\
396	0\\
397	0\\
398	0\\
399	0\\
400	0\\
401	0\\
402	0\\
403	0\\
404	0\\
405	0\\
406	0\\
407	0\\
408	0\\
409	0\\
410	0\\
411	0\\
412	0\\
413	0\\
414	0\\
415	0\\
416	0\\
417	0\\
418	0\\
419	0\\
420	0\\
421	0\\
422	0\\
423	0\\
424	0\\
425	0\\
426	0\\
427	0\\
428	0\\
429	0\\
430	0\\
431	0\\
432	0\\
433	0\\
434	0\\
435	0\\
436	0\\
437	0\\
438	0\\
439	0\\
440	0\\
441	0\\
442	0\\
443	0\\
444	0\\
445	0\\
446	0\\
447	0\\
448	0\\
449	0\\
450	0\\
451	0\\
452	0\\
453	0\\
454	0\\
455	0\\
456	0\\
457	0\\
458	0\\
459	0\\
460	0\\
461	0\\
462	0\\
463	0\\
464	0\\
465	0\\
466	0\\
467	0\\
468	0\\
469	0\\
470	0\\
471	0\\
472	0\\
473	0\\
474	0\\
475	0\\
476	0\\
477	0\\
478	0\\
479	0\\
480	0\\
481	0\\
482	0\\
483	0\\
484	0\\
485	0\\
486	0\\
487	0\\
488	0\\
489	0\\
490	0\\
491	0\\
492	0\\
493	0\\
494	0\\
495	0\\
496	0\\
497	0\\
498	0\\
499	0\\
500	0\\
501	0\\
502	0\\
503	0\\
504	0\\
505	0\\
506	0\\
507	0\\
508	0\\
509	-0.72\\
510	-0.623\\
511	-0.307\\
512	0\\
513	0\\
514	0\\
515	0\\
516	0\\
517	0\\
518	0\\
519	0\\
520	0\\
521	0\\
522	0\\
523	0\\
524	0\\
525	0\\
526	0\\
527	0\\
528	0\\
529	0\\
530	0\\
531	0\\
532	0\\
533	0\\
534	0\\
535	0\\
536	0\\
537	0\\
538	0\\
539	0\\
540	0\\
541	0\\
542	0\\
543	0\\
544	0\\
545	0\\
546	0\\
547	0\\
548	0\\
549	0\\
550	0\\
551	0\\
552	0\\
553	0\\
554	0\\
555	0\\
556	0\\
557	0\\
558	0\\
559	0\\
560	0\\
561	0\\
562	0\\
563	0\\
564	0\\
565	0\\
566	0\\
567	0\\
568	0\\
569	0\\
570	0\\
571	0\\
572	0\\
573	0\\
574	0\\
575	0\\
576	0\\
577	0\\
578	0\\
579	0\\
580	0\\
581	0\\
582	0\\
583	0\\
584	0\\
585	0\\
586	0\\
587	0\\
588	0\\
589	0\\
590	0\\
591	0\\
592	0\\
593	0\\
594	0\\
595	0\\
596	0\\
597	0\\
598	0\\
599	0\\
600	0\\
601	0\\
602	0\\
603	0\\
604	0\\
605	0\\
606	0\\
607	0\\
608	0\\
609	0\\
610	0\\
611	0\\
612	0\\
613	0\\
614	0\\
615	0\\
616	0\\
617	0\\
618	0\\
619	0\\
620	0\\
621	0\\
622	0\\
623	0\\
624	0\\
625	0\\
626	0\\
627	0\\
628	0\\
629	0\\
630	0\\
631	0\\
632	0\\
633	0\\
634	0\\
635	0\\
636	0.72\\
637	0.623\\
638	0.307\\
639	0\\
640	0\\
641	0\\
642	0\\
643	0\\
644	0\\
645	0\\
646	0\\
647	0\\
648	0\\
649	0\\
650	0\\
651	0\\
652	0\\
653	0\\
654	0\\
655	0\\
656	0\\
657	0\\
658	0\\
659	0\\
660	0\\
661	0\\
662	0\\
663	0\\
664	0\\
665	0\\
666	0\\
667	0\\
668	0\\
669	0\\
670	0\\
671	0\\
672	0\\
673	0\\
674	0\\
675	0\\
676	0\\
677	0\\
678	0\\
679	0\\
680	0\\
681	0\\
682	0\\
683	0\\
684	0\\
685	0\\
686	0\\
687	0\\
688	0\\
689	0\\
690	0\\
691	0\\
692	0\\
693	0\\
694	0\\
695	0\\
696	0\\
697	0\\
698	0\\
699	0\\
700	0\\
701	0\\
702	0\\
703	0\\
704	0\\
705	0\\
706	0\\
707	0\\
708	0\\
709	0\\
710	0\\
711	0\\
712	0\\
713	0\\
714	0\\
715	0\\
716	0\\
717	0\\
718	0\\
719	0\\
720	0\\
721	0\\
722	0\\
723	0\\
724	0\\
725	0\\
726	0\\
727	0\\
728	0\\
729	0\\
730	0\\
731	0\\
732	0\\
733	0\\
734	0\\
735	0\\
736	0\\
737	0\\
738	0\\
739	0\\
740	0\\
741	0\\
742	0\\
743	0\\
744	0\\
745	0\\
746	0\\
747	0\\
748	0\\
749	0\\
750	0\\
751	0\\
752	0\\
753	0\\
754	0\\
755	0\\
756	0\\
757	0\\
758	0\\
759	0\\
760	0\\
761	0\\
762	0\\
};
\addplot[forget plot, color=white!15!black] table[row sep=crcr] {%
0	0\\
762	0\\
};
\addlegendentry{theory, eq. (3)}

\addplot[ycomb, color=blue, mark=+, mark options={solid, blue}] table[row sep=crcr] {%
1	1.44\\
2	1.246\\
3	0.614\\
4	0\\
5	0\\
6	0\\
7	0\\
8	0\\
9	0\\
10	0\\
11	0\\
12	0\\
13	0\\
14	0\\
15	0\\
16	0\\
17	0\\
18	0\\
19	0\\
20	0\\
21	0\\
22	0\\
23	0\\
24	0\\
25	0\\
26	0\\
27	0\\
28	0\\
29	0\\
30	0\\
31	0\\
32	0\\
33	0\\
34	0\\
35	0\\
36	0\\
37	0\\
38	0\\
39	0\\
40	0\\
41	0\\
42	0\\
43	0\\
44	0\\
45	0\\
46	0\\
47	0\\
48	0\\
49	0\\
50	0\\
51	0\\
52	0\\
53	0\\
54	0\\
55	0\\
56	0\\
57	0\\
58	0\\
59	0\\
60	0\\
61	0\\
62	0\\
63	0\\
64	0\\
65	0\\
66	0\\
67	0\\
68	0\\
69	0\\
70	0\\
71	0\\
72	0\\
73	0\\
74	0\\
75	0\\
76	0\\
77	0\\
78	0\\
79	0\\
80	0\\
81	0\\
82	0\\
83	0\\
84	0\\
85	0\\
86	0\\
87	0\\
88	0\\
89	0\\
90	0\\
91	0\\
92	0\\
93	0\\
94	0\\
95	0\\
96	0\\
97	0\\
98	0\\
99	0\\
100	0\\
101	0\\
102	0\\
103	0\\
104	0\\
105	0\\
106	0\\
107	0\\
108	0\\
109	0\\
110	0\\
111	0\\
112	0\\
113	0\\
114	0\\
115	0\\
116	0\\
117	0\\
118	0\\
119	0\\
120	0\\
121	0\\
122	0\\
123	0\\
124	0\\
125	0\\
126	0\\
127	0\\
128	0.72\\
129	0.623\\
130	0.307\\
131	0\\
132	0\\
133	0\\
134	0\\
135	0\\
136	0\\
137	0\\
138	0\\
139	0\\
140	0\\
141	0\\
142	0\\
143	0\\
144	0\\
145	0\\
146	0\\
147	0\\
148	0\\
149	0\\
150	0\\
151	0\\
152	0\\
153	0\\
154	0\\
155	0\\
156	0\\
157	0\\
158	0\\
159	0\\
160	0\\
161	0\\
162	0\\
163	0\\
164	0\\
165	0\\
166	0\\
167	0\\
168	0\\
169	0\\
170	0\\
171	0\\
172	0\\
173	0\\
174	0\\
175	0\\
176	0\\
177	0\\
178	0\\
179	0\\
180	0\\
181	0\\
182	0\\
183	0\\
184	0\\
185	0\\
186	0\\
187	0\\
188	0\\
189	0\\
190	0\\
191	0\\
192	0\\
193	0\\
194	0\\
195	0\\
196	0\\
197	0\\
198	0\\
199	0\\
200	0\\
201	0\\
202	0\\
203	0\\
204	0\\
205	0\\
206	0\\
207	0\\
208	0\\
209	0\\
210	0\\
211	0\\
212	0\\
213	0\\
214	0\\
215	0\\
216	0\\
217	0\\
218	0\\
219	0\\
220	0\\
221	0\\
222	0\\
223	0\\
224	0\\
225	0\\
226	0\\
227	0\\
228	0\\
229	0\\
230	0\\
231	0\\
232	0\\
233	0\\
234	0\\
235	0\\
236	0\\
237	0\\
238	0\\
239	0\\
240	0\\
241	0\\
242	0\\
243	0\\
244	0\\
245	0\\
246	0\\
247	0\\
248	0\\
249	0\\
250	0\\
251	0\\
252	0\\
253	0\\
254	-0\\
255	-0.72\\
256	-0.623\\
257	-0.307\\
258	-0\\
259	-0\\
260	-0\\
261	-0\\
262	-0\\
263	-0\\
264	-0\\
265	-0\\
266	-0\\
267	-0\\
268	-0\\
269	-0\\
270	-0\\
271	-0\\
272	-0\\
273	-0\\
274	-0\\
275	-0\\
276	-0\\
277	-0\\
278	-0\\
279	-0\\
280	-0\\
281	-0\\
282	-0\\
283	-0\\
284	-0\\
285	-0\\
286	-0\\
287	-0\\
288	-0\\
289	-0\\
290	-0\\
291	-0\\
292	-0\\
293	-0\\
294	-0\\
295	-0\\
296	-0\\
297	-0\\
298	-0\\
299	-0\\
300	-0\\
301	-0\\
302	-0\\
303	-0\\
304	-0\\
305	-0\\
306	-0\\
307	-0\\
308	-0\\
309	-0\\
310	-0\\
311	-0\\
312	-0\\
313	-0\\
314	-0\\
315	-0\\
316	-0\\
317	-0\\
318	-0\\
319	-0\\
320	-0\\
321	-0\\
322	-0\\
323	-0\\
324	-0\\
325	-0\\
326	-0\\
327	-0\\
328	-0\\
329	-0\\
330	-0\\
331	-0\\
332	-0\\
333	-0\\
334	-0\\
335	-0\\
336	-0\\
337	-0\\
338	-0\\
339	-0\\
340	-0\\
341	-0\\
342	-0\\
343	-0\\
344	-0\\
345	-0\\
346	-0\\
347	-0\\
348	-0\\
349	-0\\
350	-0\\
351	-0\\
352	-0\\
353	-0\\
354	-0\\
355	-0\\
356	-0\\
357	-0\\
358	-0\\
359	-0\\
360	-0\\
361	-0\\
362	-0\\
363	-0\\
364	-0\\
365	-0\\
366	-0\\
367	-0\\
368	-0\\
369	-0\\
370	-0\\
371	-0\\
372	-0\\
373	-0\\
374	-0\\
375	-0\\
376	-0\\
377	-0\\
378	-0\\
379	-0\\
380	-0\\
381	-0\\
382	-1.44\\
383	-1.246\\
384	-0.614\\
385	-0\\
386	-0\\
387	-0\\
388	-0\\
389	-0\\
390	-0\\
391	-0\\
392	-0\\
393	-0\\
394	-0\\
395	-0\\
396	-0\\
397	-0\\
398	-0\\
399	-0\\
400	-0\\
401	-0\\
402	-0\\
403	-0\\
404	-0\\
405	-0\\
406	-0\\
407	-0\\
408	-0\\
409	-0\\
410	-0\\
411	-0\\
412	-0\\
413	-0\\
414	-0\\
415	-0\\
416	-0\\
417	-0\\
418	-0\\
419	-0\\
420	-0\\
421	-0\\
422	-0\\
423	-0\\
424	-0\\
425	-0\\
426	-0\\
427	-0\\
428	-0\\
429	-0\\
430	-0\\
431	-0\\
432	-0\\
433	-0\\
434	-0\\
435	-0\\
436	-0\\
437	-0\\
438	-0\\
439	-0\\
440	-0\\
441	-0\\
442	-0\\
443	-0\\
444	-0\\
445	-0\\
446	-0\\
447	-0\\
448	-0\\
449	-0\\
450	-0\\
451	-0\\
452	-0\\
453	-0\\
454	-0\\
455	-0\\
456	-0\\
457	-0\\
458	-0\\
459	-0\\
460	-0\\
461	-0\\
462	-0\\
463	-0\\
464	-0\\
465	-0\\
466	-0\\
467	-0\\
468	-0\\
469	-0\\
470	-0\\
471	-0\\
472	-0\\
473	-0\\
474	-0\\
475	-0\\
476	-0\\
477	-0\\
478	-0\\
479	-0\\
480	-0\\
481	-0\\
482	-0\\
483	-0\\
484	-0\\
485	-0\\
486	-0\\
487	-0\\
488	-0\\
489	-0\\
490	-0\\
491	-0\\
492	-0\\
493	-0\\
494	-0\\
495	-0\\
496	-0\\
497	-0\\
498	-0\\
499	-0\\
500	-0\\
501	-0\\
502	-0\\
503	-0\\
504	-0\\
505	-0\\
506	-0\\
507	-0\\
508	-0\\
509	-0.72\\
510	-0.623\\
511	-0.307\\
512	-0\\
513	-0\\
514	-0\\
515	-0\\
516	-0\\
517	-0\\
518	-0\\
519	-0\\
520	-0\\
521	-0\\
522	-0\\
523	-0\\
524	-0\\
525	-0\\
526	-0\\
527	-0\\
528	-0\\
529	-0\\
530	-0\\
531	-0\\
532	-0\\
533	-0\\
534	-0\\
535	-0\\
536	-0\\
537	-0\\
538	-0\\
539	-0\\
540	-0\\
541	-0\\
542	-0\\
543	-0\\
544	-0\\
545	-0\\
546	-0\\
547	-0\\
548	-0\\
549	-0\\
550	-0\\
551	-0\\
552	-0\\
553	-0\\
554	-0\\
555	-0\\
556	-0\\
557	-0\\
558	-0\\
559	-0\\
560	-0\\
561	-0\\
562	-0\\
563	-0\\
564	-0\\
565	-0\\
566	-0\\
567	-0\\
568	-0\\
569	-0\\
570	-0\\
571	-0\\
572	-0\\
573	-0\\
574	-0\\
575	-0\\
576	-0\\
577	-0\\
578	-0\\
579	-0\\
580	-0\\
581	-0\\
582	-0\\
583	-0\\
584	-0\\
585	-0\\
586	-0\\
587	-0\\
588	-0\\
589	-0\\
590	-0\\
591	-0\\
592	-0\\
593	-0\\
594	-0\\
595	-0\\
596	-0\\
597	-0\\
598	-0\\
599	-0\\
600	-0\\
601	-0\\
602	-0\\
603	-0\\
604	-0\\
605	-0\\
606	-0\\
607	-0\\
608	-0\\
609	-0\\
610	-0\\
611	-0\\
612	-0\\
613	-0\\
614	-0\\
615	-0\\
616	-0\\
617	-0\\
618	-0\\
619	-0\\
620	-0\\
621	-0\\
622	-0\\
623	-0\\
624	-0\\
625	-0\\
626	-0\\
627	-0\\
628	-0\\
629	-0\\
630	-0\\
631	-0\\
632	-0\\
633	-0\\
634	-0\\
635	0\\
636	0.72\\
637	0.623\\
638	0.307\\
639	0\\
640	0\\
641	0\\
642	0\\
643	0\\
644	0\\
645	0\\
646	0\\
647	0\\
648	0\\
649	0\\
650	0\\
651	0\\
652	0\\
653	0\\
654	0\\
655	0\\
656	0\\
657	0\\
658	0\\
659	0\\
660	0\\
661	0\\
662	0\\
663	0\\
664	0\\
665	0\\
666	0\\
667	0\\
668	0\\
669	0\\
670	0\\
671	0\\
672	0\\
673	0\\
674	0\\
675	0\\
676	0\\
677	0\\
678	0\\
679	0\\
680	0\\
681	0\\
682	0\\
683	0\\
684	0\\
685	0\\
686	0\\
687	0\\
688	0\\
689	0\\
690	0\\
691	0\\
692	0\\
693	0\\
694	0\\
695	0\\
696	0\\
697	0\\
698	0\\
699	0\\
700	0\\
701	0\\
702	0\\
703	0\\
704	0\\
705	0\\
706	0\\
707	0\\
708	0\\
709	0\\
710	0\\
711	0\\
712	0\\
713	0\\
714	0\\
715	0\\
716	0\\
717	0\\
718	0\\
719	0\\
720	0\\
721	0\\
722	0\\
723	0\\
724	0\\
725	0\\
726	0\\
727	0\\
728	0\\
729	0\\
730	0\\
731	0\\
732	0\\
733	0\\
734	0\\
735	0\\
736	0\\
737	0\\
738	0\\
739	0\\
740	0\\
741	0\\
742	0\\
743	0\\
744	0\\
745	0\\
746	0\\
747	0\\
748	0\\
749	0\\
750	0\\
751	0\\
752	0\\
753	0\\
754	0\\
755	0\\
756	0\\
757	0\\
758	0\\
759	0\\
760	0\\
761	0\\
762	0\\
};
\addplot[forget plot, color=white!15!black] table[row sep=crcr] {%
0	0\\
762	0\\
};
\addlegendentry{theory, eq. (4)}

\node at (720,1.3) {(a)};

\end{axis}
\end{tikzpicture}%

%% file: Fig_4_b.tex
%
%
\definecolor{mycolor1}{rgb}{0.00000,0.44700,0.74100}%
\begin{tikzpicture}

\begin{axis}[%
width=8.5cm, height=4cm,
at={(1cm,0cm)},
xmin=-0.212520868170715,
xmax=10.23708519694,
xlabel style={font=\color{white!15!black}},
xlabel={$r$},
ymin=-0.2,
ymax=1.6,
ylabel style={font=\color{white!15!black}},
ylabel={$R_{yu}(r)$},
axis background/.style={fill=white}
]
\addplot[ycomb, color=mycolor1, mark=o, mark options={solid, mycolor1}] table[row sep=crcr] {%
1	1.43767884514436\\
2	1.24129936399474\\
3	0.611368053947369\\
4	-0.00166503557312254\\
5	-0.00410822163588391\\
6	-0.00283719550858656\\
7	-0.00158350925925927\\
8	3.93668874172194e-05\\
9	-0.00067454907161803\\
10	0.00320862682602923\\
11	0.00200484308510637\\
};
\addplot[forget plot, color=white!15!black] table[row sep=crcr] {%
-0.212520868170715	0\\
10.23708519694	0\\
};

\addplot[ycomb, color=black, mark=x, mark options={solid, black}] table[row sep=crcr] {%
1	1.44\\
2	1.246\\
3	0.614\\
4	0\\
5	0\\
6	0\\
7	0\\
8	0\\
9	0\\
10	0\\
11	0\\
};
\addplot[forget plot, color=white!15!black] table[row sep=crcr] {%
-0.212520868170715	0\\
10.23708519694	0\\
};

\addplot[ycomb, color=blue, mark=+, mark options={solid, blue}] table[row sep=crcr] {%
1	1.44\\
2	1.246\\
3	0.614\\
4	0\\
5	0\\
6	0\\
7	0\\
8	0\\
9	0\\
10	0\\
11	0\\
};
\addplot[forget plot, color=white!15!black] table[row sep=crcr] {%
-0.212520868170715	0\\
10.23708519694	0\\
};

\node at (9.7, 1.4) {(b)};

\end{axis}
\end{tikzpicture}%

%% file: plot_clip_G_ratio_A1.tex
	
\begin{axis}[
grid=major, 
xmin=0, xmax=10000,	
width=8.5cm, height=4.00 cm, 
legend columns=-1,
legend to name=legend_location3,
xlabel={\footnotesize Frequency [Hz]},
ylabel={\footnotesize $\left|G/\hat{G}_{BLA}\right|$ [dB]},
clip marker paths=true
]	

	\addplot 
	[mark=*, mark size=1pt, only marks, 
		mark options={draw=blue,fill=blue, solid}] 
	table[x expr=\thisrowno{0}, y expr=\thisrowno{4}, col sep =comma] 
	{./data/data_clip_A1_1.txt};
	\addlegendentry{DS; \ \ }
	
	\addplot 
	[mark=+, mark size=1.5pt, only marks, 
		mark options={draw=black, fill=black, solid}] 
	table[x expr=\thisrowno{0}, y expr=\thisrowno{4}, col sep =comma] 
	{./data/data_clip_A1_2.txt};
	\addlegendentry{RCS;  \ \ }	
	
	\addplot 
	[mark=square, mark size=1.5pt, only marks, 
		mark options={draw=green, fill=green, solid}] 
	table[x expr=\thisrowno{0}, y expr=\thisrowno{4}, col sep =comma] 
	{./data/data_clip_A1_3.txt};
	\addlegendentry{WGN;  \ \ }	
	
	\addplot 
	[mark=x, mark size=1.5pt, only marks, 
		mark options={draw=cyan, fill=cyan, solid}] 
	table[x expr=\thisrowno{0}, y expr=\thisrowno{4}, col sep =comma] 
	{./data/data_clip_A1_4.txt};
	\addlegendentry{MLBS;  \ \ }	
	
	\addplot 
	[mark=o, mark size=1.5pt, only marks, 
		mark options={draw=magenta, fill=magenta, solid}] 
	table[x expr=\thisrowno{0}, y expr=\thisrowno{4}, col sep =comma] 
	{./data/data_clip_A1_5.txt};
	\addlegendentry{IRMLBS;  \ \ }
	
	\addplot 
	[dashed, draw=blue, line width=2pt, samples=2, domain=0:10000] {0};
	\addlegendentry{linear}		

\end{axis}
	

%% file: plot_clip_G_ratio_A2.tex
\begin{axis}[
grid=major, 
xmin=0, xmax=10000,	
width=8.5cm, height=4.00 cm, 
xlabel={\footnotesize Frequency [Hz]},
ylabel={\footnotesize $\left|G/\hat{G}_{BLA}\right|$ [dB]},
clip marker paths=true
]	

	\addplot 
	[mark=*, mark size=1pt, only marks, 
		mark options={draw=blue,fill=blue, solid}] 
	table[x expr=\thisrowno{0}, y expr=\thisrowno{4}, col sep =comma] 
	{./data/data_clip_A2_1.txt};
	
	\addplot 
	[mark=+, mark size=1.5pt, only marks, 
		mark options={draw=black, fill=black, solid}] 
	table[x expr=\thisrowno{0}, y expr=\thisrowno{4}, col sep =comma] 
	{./data/data_clip_A2_2.txt};
	
	\addplot 
	[mark=square, mark size=1.5pt, only marks, 
		mark options={draw=green, fill=green, solid}] 
	table[x expr=\thisrowno{0}, y expr=\thisrowno{4}, col sep =comma] 
	{./data/data_clip_A2_3.txt};
	
	\addplot 
	[mark=x, mark size=1.5pt, only marks, 
		mark options={draw=cyan, fill=cyan, solid}] 
	table[x expr=\thisrowno{0}, y expr=\thisrowno{4}, col sep =comma] 
	{./data/data_clip_A2_4.txt};
	
	\addplot 
	[mark=o, mark size=1.5pt, only marks, 
		mark options={draw=magenta, fill=magenta, solid}] 
	table[x expr=\thisrowno{0}, y expr=\thisrowno{4}, col sep =comma] 
	{./data/data_clip_A2_5.txt};
	
	\addplot 
	[dashed, draw=blue, line width=2pt, samples=2, domain=0:10000] {0};

\end{axis}
	

%% file: plot_clip_G_phase_diff_A1.tex
\begin{axis}[
grid=major, 
xmin=0, xmax=10000,	
width=8.5cm, height=4.00 cm, 
xlabel={\footnotesize Frequency [Hz]},
ylabel={\footnotesize $\angle\left(G\right) - \angle\left(\hat{G}_{BLA}\right)$ [deg]},
clip marker paths=true
]

	\addplot 
	[mark=*, mark size=1pt, only marks, 
		mark options={draw=blue,fill=blue, solid}] 
	table[x expr=\thisrowno{0}, y expr=\thisrowno{5}, col sep =comma] 
	{./data/data_clip_A1_1.txt};
	
	\addplot 
	[mark=+, mark size=1.5pt, only marks, 
		mark options={draw=black, fill=black, solid}] 
	table[x expr=\thisrowno{0}, y expr=\thisrowno{5}, col sep =comma] 
	{./data/data_clip_A1_2.txt};
	
	\addplot 
	[mark=square, mark size=1.5pt, only marks, 
		mark options={draw=green, fill=green, solid}] 
	table[x expr=\thisrowno{0}, y expr=\thisrowno{5}, col sep =comma] 
	{./data/data_clip_A1_3.txt};
	
	\addplot 
	[mark=x, mark size=1.5pt, only marks, 
		mark options={draw=cyan, fill=cyan, solid}] 
	table[x expr=\thisrowno{0}, y expr=\thisrowno{5}, col sep =comma] 
	{./data/data_clip_A1_4.txt};
	
	\addplot 
	[mark=o, mark size=1.5pt, only marks, 
		mark options={draw=magenta, fill=magenta, solid}] 
	table[x expr=\thisrowno{0}, y expr=\thisrowno{5}, col sep =comma] 
	{./data/data_clip_A1_5.txt};
	
	\addplot 
	[dashed, draw=blue, line width=2pt, samples=2, domain=0:10000] {0};

\end{axis}

%% file: plot_clip_G_phase_diff_A2.tex
\begin{axis}[
grid=major, 
xmin=0, xmax=10000,	
width=8.5cm, height=4.00 cm, 
legend style={draw=none},
legend columns=-1,
xlabel={\footnotesize Frequency [Hz]},
ylabel={\footnotesize $\angle\left(G\right) - \angle\left(\hat{G}_{BLA}\right)$ [deg]},
clip marker paths=true
]

	\addplot 
	[mark=*, mark size=1pt, only marks, 
		mark options={draw=blue,fill=blue, solid}] 
	table[x expr=\thisrowno{0}, y expr=\thisrowno{5}, col sep =comma] 
	{./data/data_clip_A2_1.txt};
	
	\addplot 
	[mark=+, mark size=1.5pt, only marks, 
		mark options={draw=black, fill=black, solid}] 
	table[x expr=\thisrowno{0}, y expr=\thisrowno{5}, col sep =comma] 
	{./data/data_clip_A2_2.txt};
	
	\addplot 
	[mark=square, mark size=1.5pt, only marks, 
		mark options={draw=green, fill=green, solid}] 
	table[x expr=\thisrowno{0}, y expr=\thisrowno{5}, col sep =comma] 
	{./data/data_clip_A2_3.txt};
	
	\addplot 
	[mark=x, mark size=1.5pt, only marks, 
		mark options={draw=cyan, fill=cyan, solid}] 
	table[x expr=\thisrowno{0}, y expr=\thisrowno{5}, col sep =comma] 
	{./data/data_clip_A2_4.txt};
	
	\addplot 
	[mark=o, mark size=1.5pt, only marks, 
		mark options={draw=magenta, fill=magenta, solid}] 
	table[x expr=\thisrowno{0}, y expr=\thisrowno{5}, col sep =comma] 
	{./data/data_clip_A2_5.txt};
	
	\addplot 
	[dashed, draw=blue, line width=2pt, samples=2, domain=0:10000] {0};

\end{axis}

%% file: plot_exper_clip_G_ratio_A1.tex
	
\begin{axis}[
grid=major, 
xmin=0, xmax=10000,	
width=8.5cm, height=4.00 cm, 
legend columns=-1,
legend to name=legend_location4,
xlabel={\footnotesize Frequency [Hz]},
ylabel={\footnotesize $\left|G/\hat{G}_{BLA}\right|$ [dB]},
clip marker paths=true
]	

	\addplot 
	[mark=*, mark size=1pt, only marks, 
		mark options={draw=blue,fill=blue, solid}] 
	table[x expr=\thisrowno{0}, y expr=\thisrowno{1}, col sep =comma] 
	{./data/data_exper_clip_A1_1_ratio.txt};
	\addlegendentry{DS; \ \ }
	
	\addplot 
	[mark=+, mark size=1.5pt, only marks, 
		mark options={draw=black, fill=black, solid}] 
	table[x expr=\thisrowno{0}, y expr=\thisrowno{1}, col sep =comma] 
	{./data/data_exper_clip_A1_2_ratio.txt};
	\addlegendentry{RCS;  \ \ }	
	
	\addplot 
	[mark=square, mark size=1.5pt, only marks, 
		mark options={draw=green, fill=green, solid}] 
	table[x expr=\thisrowno{0}, y expr=\thisrowno{1}, col sep =comma] 
	{./data/data_exper_clip_A1_3_ratio.txt};
	\addlegendentry{WGN;  \ \ }	
	
	\addplot 
	[mark=x, mark size=1.5pt, only marks, 
		mark options={draw=cyan, fill=cyan, solid}] 
	table[x expr=\thisrowno{0}, y expr=\thisrowno{1}, col sep =comma] 
	{./data/data_exper_clip_A1_4_ratio.txt};
	\addlegendentry{MLBS;  \ \ }	
	
	\addplot 
	[mark=o, mark size=1.5pt, only marks, 
		mark options={draw=magenta, fill=magenta, solid}] 
	table[x expr=\thisrowno{0}, y expr=\thisrowno{1}, col sep =comma] 
	{./data/data_exper_clip_A1_5_ratio.txt};
	\addlegendentry{IRMLBS;  \ \ }
	
	\addplot 
	[dashed, draw=blue, line width=2pt, samples=2, domain=0:10000] {0};
	\addlegendentry{linear (stepped sine)}		

\end{axis}
	

%% file: plot_exper_clip_G_ratio_A2.tex
	
\begin{axis}[
grid=major, 
xmin=0, xmax=10000,	
width=8.5cm, height=4.00 cm, 
legend columns=-1,
legend to name=legend_location,
xlabel={\footnotesize Frequency [Hz]},
ylabel={\footnotesize $\left|G/\hat{G}_{BLA}\right|$ [dB]},
clip marker paths=true
]	

	\addplot 
	[mark=*, mark size=1pt, only marks, 
		mark options={draw=blue,fill=blue, solid}] 
	table[x expr=\thisrowno{0}, y expr=\thisrowno{1}, col sep =comma] 
	{./data/data_exper_clip_A2_1_ratio.txt};
	\addlegendentry{DS; \ \ }
	
	\addplot 
	[mark=+, mark size=1.5pt, only marks, 
		mark options={draw=black, fill=black, solid}] 
	table[x expr=\thisrowno{0}, y expr=\thisrowno{1}, col sep =comma] 
	{./data/data_exper_clip_A2_2_ratio.txt};
	\addlegendentry{RCS;  \ \ }	
	
	\addplot 
	[mark=square, mark size=1.5pt, only marks, 
		mark options={draw=green, fill=green, solid}] 
	table[x expr=\thisrowno{0}, y expr=\thisrowno{1}, col sep =comma] 
	{./data/data_exper_clip_A2_3_ratio.txt};
	\addlegendentry{WGN;  \ \ }	
	
	\addplot 
	[mark=x, mark size=1.5pt, only marks, 
		mark options={draw=cyan, fill=cyan, solid}] 
	table[x expr=\thisrowno{0}, y expr=\thisrowno{1}, col sep =comma] 
	{./data/data_exper_clip_A2_4_ratio.txt};
	\addlegendentry{MLBS;  \ \ }	
	
	\addplot 
	[mark=o, mark size=1.5pt, only marks, 
		mark options={draw=magenta, fill=magenta, solid}] 
	table[x expr=\thisrowno{0}, y expr=\thisrowno{1}, col sep =comma] 
	{./data/data_exper_clip_A2_5_ratio.txt};
	\addlegendentry{IRMLBS;  \ \ }
	
	\addplot 
	[dashed, draw=blue, line width=2pt, samples=2, domain=0:10000] {0};
	\addlegendentry{linear (stepped sine)}		

\end{axis}
	

%% file: plot_exper_clip_G_phase_diff_A1.tex
\begin{axis}[
grid=major, 
xmin=0, xmax=10000,	
width=8.5cm, height=4.00 cm, 
xlabel={\footnotesize Frequency [Hz]},
ylabel={\footnotesize $\angle\left(G\right) - \angle\left(\hat{G}_{BLA}\right)$ [deg]},
clip marker paths=true
]

	\addplot 
	[mark=*, mark size=1pt, only marks, 
		mark options={draw=blue,fill=blue, solid}] 
	table[x expr=\thisrowno{0}, y expr=\thisrowno{2}, col sep =comma] 
	{./data/data_exper_clip_A1_1_ratio.txt};
	
	\addplot 
	[mark=+, mark size=1.5pt, only marks, 
		mark options={draw=black, fill=black, solid}] 
	table[x expr=\thisrowno{0}, y expr=\thisrowno{2}, col sep =comma] 
	{./data/data_exper_clip_A1_2_ratio.txt};
	
	\addplot 
	[mark=square, mark size=1.5pt, only marks, 
		mark options={draw=green, fill=green, solid}] 
	table[x expr=\thisrowno{0}, y expr=\thisrowno{2}, col sep =comma] 
	{./data/data_exper_clip_A1_3_ratio.txt};
	
	\addplot 
	[mark=x, mark size=1.5pt, only marks, 
		mark options={draw=cyan, fill=cyan, solid}] 
	table[x expr=\thisrowno{0}, y expr=\thisrowno{2}, col sep =comma] 
	{./data/data_exper_clip_A1_4_ratio.txt};
	
	\addplot 
	[mark=o, mark size=1.5pt, only marks, 
		mark options={draw=magenta, fill=magenta, solid}] 
	table[x expr=\thisrowno{0}, y expr=\thisrowno{2}, col sep =comma] 
	{./data/data_exper_clip_A1_5_ratio.txt};
	
	\addplot 
	[dashed, draw=blue, line width=2pt, samples=2, domain=0:10000] {0};

\end{axis}

%% file: plot_exper_clip_G_phase_diff_A2.tex
\begin{axis}[
grid=major, 
xmin=0, xmax=10000,	
width=8.5cm, height=4.00 cm, 
xlabel={\footnotesize Frequency [Hz]},
ylabel={\footnotesize $\angle\left(G\right) - \angle\left(\hat{G}_{BLA}\right)$ [deg]},
clip marker paths=true
]

	\addplot 
	[mark=*, mark size=1pt, only marks, 
		mark options={draw=blue,fill=blue, solid}] 
	table[x expr=\thisrowno{0}, y expr=\thisrowno{2}, col sep =comma] 
	{./data/data_exper_clip_A2_1_ratio.txt};
	
	\addplot 
	[mark=+, mark size=1.5pt, only marks, 
		mark options={draw=black, fill=black, solid}] 
	table[x expr=\thisrowno{0}, y expr=\thisrowno{2}, col sep =comma] 
	{./data/data_exper_clip_A2_2_ratio.txt};
	
	\addplot 
	[mark=square, mark size=1.5pt, only marks, 
		mark options={draw=green, fill=green, solid}] 
	table[x expr=\thisrowno{0}, y expr=\thisrowno{2}, col sep =comma] 
	{./data/data_exper_clip_A2_3_ratio.txt};
	
	\addplot 
	[mark=x, mark size=1.5pt, only marks, 
		mark options={draw=cyan, fill=cyan, solid}] 
	table[x expr=\thisrowno{0}, y expr=\thisrowno{2}, col sep =comma] 
	{./data/data_exper_clip_A2_4_ratio.txt};
	
	\addplot 
	[mark=o, mark size=1.5pt, only marks, 
		mark options={draw=magenta, fill=magenta, solid}] 
	table[x expr=\thisrowno{0}, y expr=\thisrowno{2}, col sep =comma] 
	{./data/data_exper_clip_A2_5_ratio.txt};
	
	\addplot 
	[dashed, draw=blue, line width=2pt, samples=2, domain=0:10000] {0};

\end{axis}

%% file: ms.bbl
\begin{thebibliography}{123}

\bibitem{Evans&Rees2000_1} C. Evans and D. Rees, ``Nonlinear distortions and multisine signals. I. Measuring the best linear approximation,'' in \emph{IEEE Trans. Instrum. Meas.}, vol. 49, no. 3, pp. 602-609, Jun 2000.

\bibitem{Ljung2001} L. Ljung, ``Estimating linear time-invariant models of nonlinear time-varying systems.'' \emph{Eur. J. Control} vol. 7, no. 2, pp. 203-219, 2001.

\bibitem{Makila2006} P.M. M\"akil\"a, ``LTI approximation of nonlinear systems via signal distribution theory,'' \emph{Automatica}, vol. 42, no. 6, pp. 917-928, June 2006.

\bibitem{EsfahaniEtAl2016} A. Fakhrizadeh Esfahani, J. Schoukens and L. Vanbeylen, ``Using the Best Linear Approximation With Varying Excitation Signals for Nonlinear System Characterization,'' in \emph{IEEE Trans. Instrum. Meas.}, vol. 65, no. 5, pp. 1271-1280, May 2016.

\bibitem{WongEtAl2012} H. K. Wong, J. Schoukens and K. R. Godfrey, ``Analysis of Best Linear Approximation of a Wiener-–Hammerstein System for Arbitrary Amplitude Distributions," in \emph{IEEE Trans. Instrum. Meas.}, vol. 61, no. 3, pp. 645-654, March 2012.

\bibitem{WongEtAl2013} H. K. Wong, J. Schoukens and K. R. Godfrey, ``Design of Multilevel Signals for Identifying the Best Linear Approximation of Nonlinear Systems," in \emph{IEEE Trans. Instrum. Meas.}, vol. 62, no. 2, pp. 519-524, Feb. 2013.

\bibitem{Tan2013} A. H. Tan, ``Direct synthesis of pseudo-random ternary perturbation signals with harmonic multiples of two and three suppressed," \emph{Automatica}, vol. 49, no. 10, pp. 2975-2981, October 2013.

\bibitem{DeAngelisEtAl2016} A. De Angelis, J. Schoukens, K. R. Godfrey and P. Carbone, ``Practical synthesis of ternary sequences for system identification,'' \emph{IEEE Int. Instrum. Meas. Technology Conference Proceedings}, pp. 423-428, Taipei, May 2016.

\bibitem{DeAngelisEtAl2016_TIM} A. De Angelis, J. Schoukens, K. R. Godfrey and P. Carbone, "Practical Issues in the Synthesis of Ternary Sequences," in \emph{IEEE Trans. Instrum. Meas.}, vol. 66, no. 2, pp. 212-222, Feb. 2017.

\balance

\bibitem{Enqvist&Ljung2005} M. Enqvist and L. Ljung, ``Linear approximations of nonlinear FIR systems for separable input processes,'' \emph{Automatica}, vol. 41, no. 3, pp. 459-473, March 2005.

\bibitem{DeAngelisEtAl2017I2MTC} A. De Angelis, J. Schoukens, K. R. Godfrey and P. Carbone, ``Measuring the Best Linear Approximation of Wiener Systems Using Multilevel Sequences'', \emph{IEEE Int. Instrum. Meas. Technology Conference Proceedings}, Torino, May 2017.

\bibitem{Pintelon&Schoukens2012} R. Pintelon and J. Schoukens, \emph{System identification: a frequency domain approach}, John Wiley and Sons, Inc., Hoboken, NJ, 2$^{nd}$ edition, 2012.

\bibitem{Evans&Rees2000_2} C. Evans and D. Rees, ``Nonlinear distortions and multisine signals. II. Minimizing the distortion,'' in \emph{IEEE Trans. Instrum. Meas.}, vol. 49, no. 3, pp. 610-616, Jun 2000.

\bibitem{Barker&Godfrey1999} H.A. Barker and K.R. Godfrey, ``System identification with multi-level periodic perturbation signals,'' \emph{Control Engineering Practice}, vol. 7, no. 6, pp. 717-726, 1999.

\bibitem{SchoukensEtAl2016} J. Schoukens, M. Vaes and R. Pintelon, ``Linear System Identification in a Nonlinear Setting: Nonparametric Analysis of the Nonlinear Distortions and Their Impact on the Best Linear Approximation,'' in \emph{IEEE Control Systems}, vol. 36, no. 3, pp. 38-69, June 2016.

\bibitem{GodfreyEtAl2005} K.R. Godfrey, A.H. Tan, H.A. Barker and B. Chong, ``A survey of readily accessible perturbation signals for system identification in the frequency domain,'' \emph{Control Engineering Practice}, vol. 13, no. 11, November 2005, pp. 1391-1402.

\bibitem{TanEtAl2005} A. H. Tan, K. R. Godfrey and H. A. Barker, ``Design of computer-optimized pseudorandom maximum length signals for linear identification in the presence of nonlinear distortions,'' in \emph{IEEE Transactions on Instrumentation and Measurement}, vol. 54, no. 6, pp. 2513-2519, Dec. 2005.

\bibitem{NewWaveInstruments} New Wave Instruments (2010): \emph{Linear feedback shift registers - implementation, m-sequence properties, feedback tables}. [Online]. Available: \url{http://www.newwaveinstruments.com/resources/articles/m_sequence_linear_feedback_shift_register_lfsr.htm}

\bibitem{Kullback&Leibler1951} S. Kullback and R. A. Leibler, ``On information and sufficiency,'' \emph{The Annals of Mathematical Statistics}, 22(1), pp. 79-86, 1951.

\bibitem{Digilent} Digilent Inc., ``Analog Discovery 2 Reference Manual,'' [Online]. Available:
\url{https://reference.digilentinc.com/reference/instrumentation/analog-discovery-2/reference-manual?redirect=1}.
Retrieved: April 2017.

\bibitem{DLMF} \emph{NIST Digital Library of Mathematical Functions, Section 26.4}. \url{http://dlmf.nist.gov/26.4}, Release 1.0.14 of 2016-12-21. F. W. J. Olver, A. B. Olde Daalhuis, D. W. Lozier, B. I. Schneider, R. F. Boisvert, C. W. Clark, B. R. Miller and B. V. Saunders, eds.






\end{thebibliography}
